\documentclass[letter,longauth]{aa}  
\usepackage{graphicx}
\usepackage{txfonts}
\usepackage[colorlinks=true,linkcolor=blue,citecolor=blue]{hyperref}
\usepackage[toc,page]{appendix}
\usepackage{float}
\usepackage{subfloat}
\usepackage{subcaption}
\usepackage[flushleft]{threeparttable}
\usepackage{overpic}
\usepackage{multirow}
\usepackage{cellspace} 
\usepackage{rotating}
\usepackage{bigstrut}
\usepackage{booktabs}

\begin{document}

   \title{SVOM GRB\,250314A at {\slshape z} $\simeq 7.3$:\\an exploding star in the era of reionization}

   \subtitle{}

   \author{B.~Cordier\inst{1}\fnmsep\thanks{E-mail: bertrand.cordier@cea.fr},
   J.~Y.~Wei\inst{2,3}\fnmsep\thanks{E-mail: wjy@nao.cas.cn},
   N.~R.~Tanvir\inst{4}\fnmsep\thanks{E-mail: nrt3@leicester.ac.uk},
   S.~D.~Vergani\inst{5,6},
   D.~B.~Malesani\inst{7,8,9},
   J.~P.~U.~Fynbo\inst{7,8},
   A.~de~Ugarte~Postigo\inst{10},
   A.~Saccardi\inst{11},
   F.~Daigne\inst{6},
   J.-L.~Atteia\inst{12},
   O.~Godet\inst{12},
   D.~G\"otz\inst{11},
   Y.~L. Qiu\inst{2},
   S.~Schanne\inst{1},
   L.~P.~Xin\inst{2},
   B.~Zhang\inst{13,14},
   S.~N.~Zhang\inst{15},
   A.~J.~Nayana\inst{16},
   L.~Piro\inst{17},
   B.~Schneider\inst{10},
   A.~J.~Levan\inst{9,18},
   A.~L.~Thakur\inst{17},
   Z.~P.~Zhu\inst{2},
   G.~Corcoran\inst{19},
   N.~A.~Rakotondrainibe\inst{10},
   V.~D'Elia\inst{20},
   D.~Turpin\inst{11}\,et~al.}

   \institute{
   CEA Paris-Saclay, Irfu/D\'epartement d’Astrophysique, 9111 Gif sur Yvette, France
   \and
   National Astronomical Observatories, Chinese Academy of Sciences, Beijing 100101, China
   \and
   School of Astronomy and Space Science, University of Chinese Academy of Sciences, Beijing 101408, China
   \and
   School of Physics and Astronomy, University of Leicester, University Road, Leicester, LE1 7RH, United Kingdom
   \and
   LUX, Observatoire de Paris, Universit\'e PSL, CNRS, Sorbonne
   Universit\'e, Meudon, 92190, France
   \and
   Sorbonne Universit\'e, CNRS, UMR 7095, Institut d'Astrophysique de Paris, 98 bis bd Arago, F-75014 Paris, France
   \and
   Niels Bohr Institute, University of Copenhagen, Jagtvej 155, 2200, Copenhagen N, Denmark
   \and
   The Cosmic Dawn Centre (DAWN), Denmark
   \and
   Department of Astrophysics/IMAPP, Radboud University, PO Box 9010, 6500 GL, The Netherlands
   \and
   Aix Marseille Univ., CNRS, CNES, LAM, Marseille, France
   \and
   Universit\'e Paris-Saclay, Universit\'e Paris Cit\'e, CEA, CNRS, AIM, 91191, Gif-sur-Yvette, France
   \and
   IRAP, Universit\'e de Toulouse, CNRS, CNES, Toulouse, France
   \and
   Department of Physics, University of Hong Kong, Pokfulam Road, Hong Kong, China
   \and
   Nevada Center for Astrophysics and Department of Physics and Astronomy, University of Nevada, Las Vegas, NV 89154, USA
   \and
   Key Laboratory of Particle Astrophysics, Institute of High Energy Physics, Chinese Academy of
   Sciences, Beijing 100049, China
   \and
   Department of Astronomy, University of California, Berkeley, CA 94720-3411, USA
   \and
   INAF, Istituto di Astrofisica e Planetologia Spaziali, Via Fosso del Cavaliere 100, Roma, 00133, Italy
   \and
   Department of Physics, University of Warwick, Coventry, CV4 7AL, UK
   \and
   School of Physics and Centre for Space Research, University College Dublin, Belfield, Dublin, D04 V1W8, Ireland
   \and
   Space Science Data Center (SSDC), Agenzia Spaziale Italiana (ASI), Roma, I-00133, Italy
}

   \date{Received September XX, YYYY; accepted March XX, YYYYY}

  \abstract
  {Most long Gamma-ray bursts (LGRBs) originate from a rare type of massive stellar explosion. Their  afterglows, while rapidly fading, can be initially extremely luminous at optical/near-infrared wavelengths, making them detectable at large cosmological distances. Here we report the detection and observations of GRB\,250314A by the \textit{SVOM} satellite and the subsequent follow-up campaign with the near-infrared afterglow discovery and the spectroscopic measurements of its redshift $z\simeq 7.3$. This burst happened when the Universe was only $\sim5$\% of its current age. We discuss the signature of these rare events within the context of the \textit{SVOM} operating model, and the ways to optimize their identification with adapted ground follow-up observation strategies.}


   \titlerunning{SVOM GRB\,250314A at {\slshape z} $\simeq 7.3$}
   \authorrunning{B. Cordier, J. Y. Wei, N. R. Tanvir et al.}

   \maketitle
%

\section{Introduction}
Long Gamma-ray bursts have long been regarded as powerful tools to explore the early Universe. Originating from the explosion of rare massive stars \citep{Hjorth2003,Woosley2006a,Hjorth2012,Cano2017}, they produce intense afterglow emission that can be detected up to the highest redshifts, e.g. $z\sim10$, and beyond \citep{Kann2024}. Their direct association with individual stars makes them key tracers of star formation \citep[e.g.][]{Krogager2024}, including when their host galaxies are too faint to be observed directly through emission lines, even with sensitive facilities such as the {\it{James Webb} Space Telescope} ({\it JWST}). With multiple GRBs detected at $z>6$, we could begin probing the high-redshift Universe and chemically characterize their host interstellar medium (ISM) thanks to GRB afterglow spectroscopy \citep{Hartoog2015,Saccardi23,Saccardi25}. Furthermore, precise analysis of the Lyman-$\alpha$ break could allow an estimation of the neutral hydrogen column density of the GRB host galaxy \citep{Tanvir2019}. The new identification of distant events is therefore highly desirable. However, to date, we are statistically limited by a small number of GRBs at very high redshift ($z>7$), and only two have a well-constrained spectroscopic redshift, with a low signal-to-noise spectrum without identification of metal absorption lines, $z_{spec}=8.23$ \citep[GRB 090423A;][]{Salvaterra2009,Tanvir2009} and $z_{spec}=7.8$ \citep[GRB\,120923A;][]{Tanvir2018}. Other high-redshift GRBs have been discovered, but only a photometric redshift has been derived: $z_{phot}\simeq9.4$ \citep[GRB 090429B;][]{Cucchiara2011} and $z_{phot}\simeq7.88$ \citep[GRB\,100905A;][]{Bolmer2018}. Given the rarity of these high-redshift events and the faintness of such sources, the scientific outcome is strongly dependent on the responsiveness of the follow-up activities. 

The new satellite \textit{SVOM} \citep[Space-based multi-band astronomical Variable Objects Monitor;][]{Wei2016}, a Sino-French mission of CNSA (China National Space Administration) and CNES (Centre National d'Études Spatiales) launched on 22 June, 2024, has been designed to favor the detection and characterization of high-redshift GRBs \citep{Cordier2008}. For this purpose,  the \textit{SVOM} collaboration has developed dedicated partnerships with other space- and ground-based facilities. In this letter we present the first result of this effort: the \textit{SVOM} detection of GRB\,250314A at $z\simeq 7.3$
and follow-up of its afterglow.

In a companion paper, \citet{levan25}
report \textit{JWST} observations that likely provide direct evidence for the association of this GRB with a massive star progenitor. 

All errors are given at the 68\% confidence level when not stated otherwise.

\vspace*{-2ex}
\section{Observations, Data analysis and Results}
\label{sec:obs}

\textit{Detection, localization and slew.}
On 2025 March 14, the \textit{SVOM}/ECLAIRs coded-mask telescope \citep{Godet2014} automatically detected and localized GRB~250314A starting at 12:56:42 UTC (Tb), used hereafter as the reference of time. Four alert packets were produced and transmitted immediately to the ground thanks to the \textit{SVOM} very high frequency (VHF) network (see Appendix~\ref{appendix:vhf_xband}). As detailed in Appendix~\ref{appendix:trigger}, the first alert reports an excess detected over an interval of 10.24\,s in the 8-50\,keV energy band starting at
Tb. The localization of the best alert 
was at $\mbox{R.A.} = 201.272^\circ$, $\mbox{Dec} =-5.293^\circ$ (J2000) with a 90\% c.l. radius of 8.62 arcmin, including 2 arcmin systematic error in quadrature \citep{GCN_SVOM}. 
The trigger requested the spacecraft slew at Tb+43~s, effectively locating GRB~250314A in the center of the field-of-view of the four instruments on-board \textit{SVOM}.

\begin{figure}[t]
    \centering
    \includegraphics*
    [viewport=0cm 0.2cm 15cm 10.9cm,width=0.9\linewidth]
    {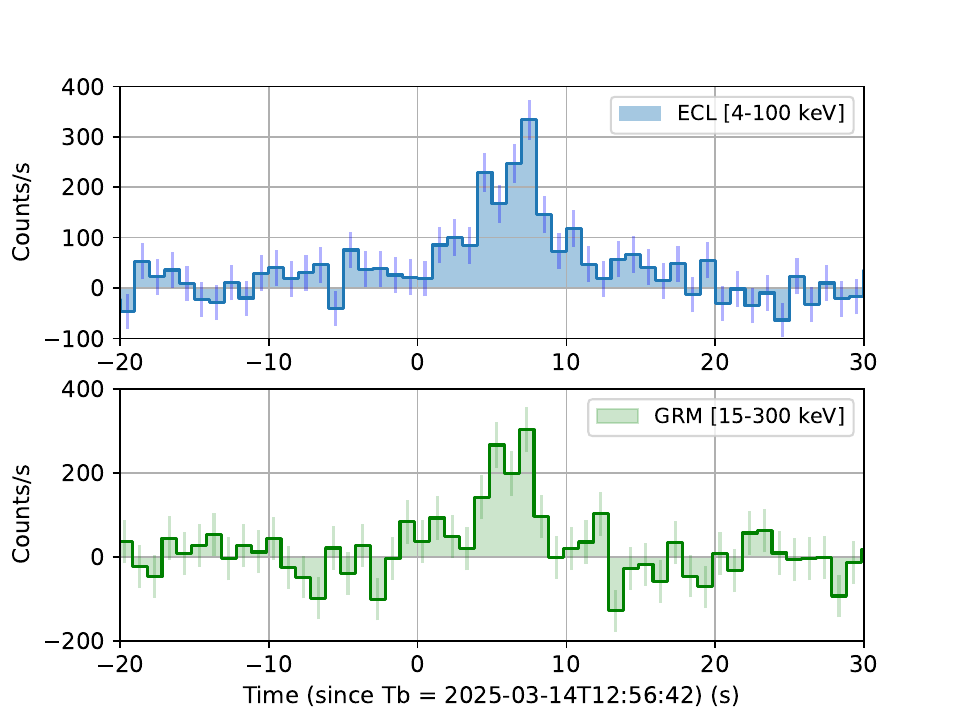}
    
    \caption{background subtracted light curves for ECLAIRs in the 4--100 keV energy range (top panel) and for GRM in the 15--300 keV energy range (bottom panel) using a time bin of 1 s.}
    \label{fig:lightcurve}
\end{figure}

\textit{Prompt GRB observations.}
GRB~250314A was detected by both gamma-ray instruments on board \textit{SVOM}, ECLAIRs and the Gamma Ray Monitor \citep[GRM,][]{GRM_intro,GRM_trigger}. 
Their data analysis is detailed in Appendix~\ref{appendix:prompt}.  
In the lightcurves shown in Figure~\ref{fig:lightcurve}, GRB~250314A appears as a weak single pulse burst, with a T90 duration of $11^{+3}_{-2}$\,s in  4--100 keV (ECLAIRs) and 7.50$^{+15.80}_{-5.00}$\,s in  15-5000 keV (GRM).
The analysis of the spectrum leads to a time-averaged flux of 5.1$^{+0.5}_{-0.9}\times10^{-8}{\rm erg\,cm^{-2}\,s^{-1}}$ in  4-120 keV (ECLAIRs)
and  3.27$^{+0.56}_{-0.55}\times10^{-8}{\rm erg\,cm^{-2}\,s^{-1}}$ in  15-5000 keV (GRM). The joint spectral analysis of ECLAIRs+GRM data shows that the time-averaged spectrum is best fitted by a powerlaw with an exponential cutoff (Fig.~\ref{fig:spectral_parameters}), with a photon index $\alpha = -1.05^{+0.22}_{-0.24}$ and a peak energy ${E}_\mathrm{p}=77^{+25}_{-14}$ keV.

\textit{X-ray follow-up with \textit{SVOM}/MXT, \textit{Swift}/XRT and \textit{EP/FXT}: afterglow detection.}
Following the slew of the  satellite, the Microchannel X-ray Telescope (MXT) \citep{mxt} on board \textit{SVOM} started observing the field of GRB\,250314A at Tb+177\,s. The MXT observation lasted until 16:18:49 UT accumulating an effective exposure time of 4950 s on source, due to Earth occultations, South Atlantic Anomaly (SAA) passages, and stray-light limitations.
The analysis of MXT data is detailed in appendix~\ref{appendix:xrays}. The afterglow of GRB~250314A is not detected. 
The first ks of
observation leads to a $3\sigma$ upper limit of 2.5$\times$10$^{-11}$ erg cm$^{-2}$ s$^{-1}$ in the 0.3-10 keV energy band.
To complement the MXT observations, we requested several Target-of-Opportunity (ToO) follow-up observations using both the \textit{Neil Gehrels Swift Observatory} \textit{Swift}/XRT \citep{Burrows2005a}  and the \textit{Einstein Probe}/FXT \citep{Chen2020} X-ray telescopes 
(0.3 -- 10 keV). As reported in Appendix~\ref{appendix:xrt_fxt},
an uncatalogued fading X-ray source in the first \textit{Swift}/XRT \citep{GCN_Swift} and \textit{EP}/FXT \citep{GCN_EP} observations is clearly detected at an enhanced \textit{Swift}/XRT position $\mbox{R.A.} = 13h 25m 12.33s$ and $\mbox{Dec} = -05^\circ 16^\prime56.1^{\prime\prime} $ (J2000) with an uncertainty of 3.4$^\prime{^\prime}$. 
As this new X-ray transient
has a position consistent with the ECLAIRs GRB one and shows a clear fading signature, we confidently identify it as the X-ray afterglow of GRB\,250314A.
The spectral analysis of \textit{Swift}/XRT and \textit{EP}/FXT data is detailed in appendix~\ref{appendix:xrays}.
In all cases the best spectral fit is obtained by an absorbed power law (photon index $\sim$ 2.4) with an $N_{H}$ fitted result which is compatible with values in excess of the Galactic one (2.46$\times$10$^{20}$ cm$^{-2}$, \citealt{HI4PI2016}): see discussion in appendix~\ref{appendix:xrays}.
The derived fluxes for all our observations are given in Tab.~ \ref{tab:obs_Xray}, and lead to the light curve shown in the top panel of Fig. \ref{fig:afterglow_lightcurve}, where it is compared to a sample of other high-z afterglows.
We notice that the flaring activity shown by several high-$z$ afterglows detected in the past is clearly excluded by our early time MXT observations.

\textit{Follow-up with \textit{SVOM}/VT.}
Following the slew, the Visible Telescope \citep[VT,][]{2020SPIE11443E..0QF} on-board \textit{SVOM} began the follow-up of the field at Tb+186\,s in VT\_B (400--650 nm) and VT\_R (650--1000 nm) channels simultaneously. The follow-up lasted for 3 orbits. 
As detailed in Appendix~\ref{appendix:vt}, no uncatalogued sources brighter than 20 mag were found within the ECLAIRs error region from 
the quicklook analysis of VHF data \citep{GCN_VT_VHF}. The stacking of the complete image data set received later revealed no candidate within 
the \textit{Swift}/XRT error box and brought to a 3$\sigma$ upper limits of 23.5 mag in VT\_B and 23.0 mag in VT\_R band at 13.9 minutes after Tb (see Tab.~\ref{VT_upperlimit}).
These early VT deep optical upper limits suggested a potential high-redshift ($z>6$) candidate \citep{GCN_VT}.

\begin{figure*}
    \centering
    \includegraphics*[viewport=0cm 0cm 67.5cm 23cm, width = 0.87\linewidth] {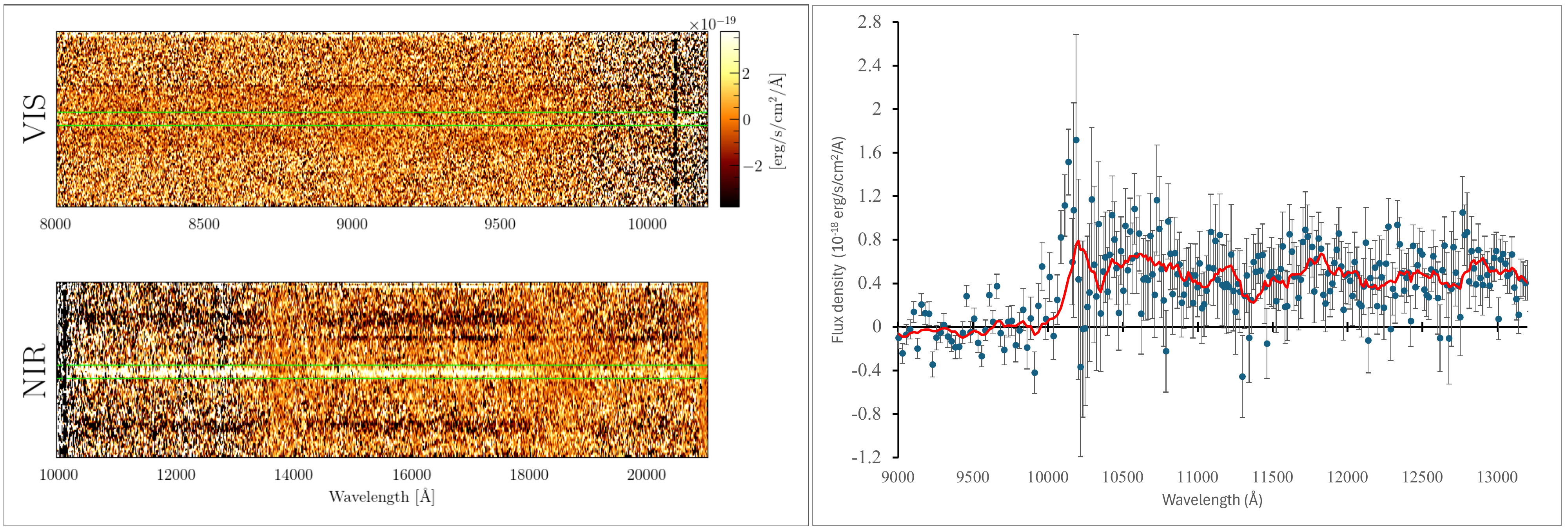}
    \caption{{\it {Left:}} 2D VIS (top) and NIR (bottom) \textit{VLT}/X-shooter spectra. The VIS spectrum is cut at 8000\,\AA\,and the vertical dashed black line mark the position corresponding to the Ly-$\alpha$ observer frame wavelength at $z\simeq 7.3$. {\it {Right:}} 1D rebinned combined spectrum from 9000 to 13000 \AA. The red solid line represents the running mean flux density along the wavelength axis. Note that the increased noise around the break location is an unfortunate consequence of it being close to the boundary between the VIS and NIR arms.}
    \label{fig:VLT_spectrum}
\end{figure*}

\textit{Ground-based follow-up observations with the NOT, the \textit{VLT} and the GTC: afterglow detection and redshift measurement.} We triggered the Nordic Optical Telescope (NOT) as soon as the X-ray localization became available. Imaging in the $J$ band was secured with the near-infrared (NIR) camera NOTCam (FoV: $4\arcmin \times 4\arcmin$), beginning 12.1~hr after Tb.
Consistent with the \textit{Swift}/XRT and \textit{EP}/FXT X-ray localisation, a single object was detected, not visible in archival images of this field, at 
$\mbox{R.A.} = 13h 25m 12.16s$ and $\mbox{Dec} = -05^\circ 16^\prime55.1^{\prime\prime} $ with an uncertainty of $0.1\arcsec$ \citep{GCN_NOT}. A second NOTCam observation was performed the following night, only yielding an upper limit to the target brightness. 
Following the identification of the counterpart, we initiated within the ``Stargate'' program photometric and spectroscopic follow-up at the ESO Very Large Telescope (\textit{VLT}), using the High Acuity Wide field K-band Imager (HAWK-I) in the $YJH$ filters, and with the X-shooter spectrograph \citep{Vernet2011}. HAWK-I observations began 16.45~hr after Tb. 
The afterglow was detected in all of the filters, clearly fading compared to the NOT measurement, thus identifying this object as the afterglow of GRB\,250314A. Photometry was also secured 16.81 hr after Tb at the 10.4-m Gran Telescopio Canarias (GTC) telescope, using the OSIRIS+ instrument and the $z$ filter. This served as ``veto'' filter, expecting a deep non-detection in the optical band due to the source spectrum drop-out. The data reduction is detailed in Appendix~\ref{appendix:v_nir} and all magnitudes are reported in Tab.~\ref{tab_photometry}, leading to the lightcurve in the bottom panel of Fig.~\ref{fig:afterglow_lightcurve}, where it is compared to the available sample of high-$z$ GRBs. 

X-shooter observations started 16.62 hr after Tb. The spectra cover the wavelength range 3000--21000~\AA\ and the spectral analysis is detailed in Appendix~\ref{appendix:v_nir}.
The spectrum shown in Fig.~\ref{fig:VLT_spectrum} (left)
reveals a flat spectral continuum that sharply drops off below $\lambda\sim$10090~\AA, consistent with a Lyman-$\alpha$ break caused by absorption from neutral hydrogen as shown in Fig.~\ref{fig:VLT_spectrum} (right). This feature indicates a redshift of $z\simeq 7.3$. Due to the very low signal-to-noise, no individual metal absorption features have been confidentially identified (see the discussion on equivalent width (EW) upper limits in Appendix~\ref{appendix:v_nir}).

\textit{Radio follow-up campaign.}
The determination of the redshift triggered a radio follow-up campaign.
The radio afterglow is detected by VLA at 6.7 days \citep{GCN_VLA} and 
ATCA at 8.96 days. ALMA observations at 11.6 days after Tb led to the most precise localization \citep{GCN_ALMA}. Data acquired with  ATCA, e-MERLIN, MeerKAT, and VLA from 6.7 to 109 days after Tb are presented and discussed in Appendix~\ref{appendix:radio}. 

\section{Discussion and Conclusions}
Our observations, triggered by the detection of GRB250314A on-board \textit{SVOM}
(timeline discussed in Appendix~\ref{appendix:timeline}), led to the identification of a new GRB above $z=7$. We summarize below the results leading us to associate this GRB to a massive star progenitor, complementary to the possible direct evidence provided by \textit{JWST} observations presented in \citet{levan25}. This confirms the importance of GRBs to probe the star formation in the early universe and leads us to discuss how the follow-up strategy may still be optimized to enlarge the high-$z$ GRB sample.

\vspace{-3ex}

\subsection{Classification of GRB\,250314A.}
\label{sec:classification}
\vspace{-1ex}
As discussed in Appendix~\ref{appendix:energetics}, the rest-frame peak energy ${E}_\mathrm{p,rest}=642^{+209}_{-118}$ keV and isotropic-equivalent energy $E_\mathrm{iso}=4.65^{+1.13}_{-0.49}\times 10^{52}\, \mathrm{erg}$ in the 10 keV--10 MeV energy range of GRB\,250314A are consistent with a classification as a ``long'' GRB (type II): see Fig.~\ref{fig:relation}.
The rather short T90 duration in the burst rest frame ($T_{90}/(1+z)=1.3^{+0.4}_{-0.2}\, \mathrm{s}$) is probably due to a ``tip of the iceberg effect'' considering the weakness of this GRB \citep{llamas24}, which is also often observed in other high redshift GRBs (see discussion in Appendix~\ref{appendix:prompt_T90}). 
As seen in Fig.~\ref{fig:relation}, there is nothing extreme in GRB\,250314A properties, even if its isotropic equivalent energy is in the bright half of the sample, as for other high redshift GRBs. We conclude that GRB\,250314A appears as a classical long (type II) GRB most probably associated to a massive star progenitor 
\citep[see e.g. discussion in][]{li:20}. 
The low-energy threshold of ECLAIRs at 4 keV favours the detection of such high-$z$ GRBs \citep{palmerio:21}.
However their early identification  remains a challenge and requires a specific strategy.

\vspace{-3ex}
\subsection{Optimizing \textit{SVOM} follow-up strategy for high-z GRBs.}
\vspace{-1ex}
\label{sec:discussion}
An important goal of the \textit{SVOM} mission is to increase the fraction of GRBs having a redshift determination, also with the aim of increasing the low number of 
high-$z$ GRBs identified to date (12 GRBs at $z>6$) and to use them as probes of stars and galaxies in the early Universe up to the reionisation era.
To reach these goals,  two key factors are necessary: a rapid determination of the precise afterglow position, and the availability of sensitive optical-NIR spectrographs. To this purpose, \textit{SVOM}  benefits from two major assets,
the slew capability, allowing the fast repointing ($<2\, \mathrm{min}$) of the on-board instruments MXT and VT, and the 
VT sensitivity and spectral coverage including the reddest part of the optical  domain.
Furthermore, for the follow-up of GRB afterglows, the \textit{SVOM} mission 
also includes dedicated ground-based facilities to perform real-time photometry,
and the \textit{SVOM} team has made agreements with the \textit{Swift} and \textit{EP} teams to obtain additional rapid X-ray follow-up observations, as well as with several collaborations having access to ground-based large telescopes to obtain deep photometry and spectroscopy.

From launch to July 12, 2025, the VT  automatically followed up 26 ECLAIRs-triggered GRBs\footnote{This represents only $\sim 60\%$ of ECLAIRs-triggered GRBs as the automatic slew was only gradually enabled after a few months.}, with a delay of several minutes after the trigger, 
and detected the optical counterpart of 19 of these (73\%). By performing the rapid analysis of VHF data (Appendix~\ref{appendix:vt}),
the first results with bright candidates can be distributed within an hour.
More precise and deeper detection results can be obtained with X-band data,
typically obtained within 5–7 hours (Appendix~\ref{appendix:vhf_xband}).
The precise localizations of the X-ray afterglow 
help us to make more firm conclusions especially in the case of no optical detection. This has already led to 14 redshift determinations (54\%).
Of the seven GRBs without optical afterglow detection, one was due to the strong stray light from Earth during the early phase, which limited the detection ability \citep{2025GCN.38824....1S}.
Three of them may be caused by heavy extinction, as indicated by the relatively high excess-column density deduced from the X-ray afterglow spectrum
\citep{2025GCN.40579....1E, 2025GCN.38918....1O}. 
GRB~250314A is one of the remaining three. 
The other two (GRB~250507A and GRB~250127A) are also high-$z$ GRB candidates, even though high-extinction or optically faint GRBs can not be excluded.

The follow-up of very high-$z$ ($z>6$) GRB candidates represents an additional challenge. There are few NIR facilities capable of ToO observations soon after a GRB detection.
As they have a limited time allocated for afterglow observations, preliminary indications of a possible high-$z$ GRB origin are necessary to activate such observations. Hence, having early time VT deep limits of the afterglow magnitude is extremely important, and the X-ray afterglow position with an accuracy better than 10{$^\prime{^\prime}$} is essential to reduce the ECLAIRs or MXT error box filtering out potential contaminants.
Both factors, played a key role in the discovery of GRB\,250314A.
This GRB also showed the importance of setting up efficient coordination and communication between \textit{SVOM} and the follow-up collaborations. The timeline in Fig.~\ref{fig:timeline} shows that there is a period of $\sim 17$ hours between the ECLAIRs trigger and the start of the \textit{VLT} observation. This delay is due to the  availability of two key pieces of information: the X-ray location and 
the VT deep upper limits.
They arrived late, respectively~11 hours and~15 hours after the trigger, whereas the corresponding observations were completed much earlier. 
The delay in determining the X-ray localization is notably long, due to data transmission gaps.
The usual delay is of $\sim3$ hours. The delay in determining deep limits from the VT data is clearly a weak point on which progress is needed.

As a conclusion, the observations of GRB~250314A presented in this letter allow us to include one more GRB in the small sample at $z>7$ and to identify the key factors that still limit the identification of such high-$z$ GRBs. 
In the future, we stress that a game-changer would be a GRB mission capable of autonomously localizing and performing spectroscopy of high-$z$ GRBs. 
In the short term, significant progress is still expected with \textit{SVOM}: 
(i) to make VT deep optical early-time upper limits available earlier than for GRB\,250314A, the main goal will be to increase the number of X-band stations to recover the full telemetry more rapidly. Once the precise X-ray localization and the VT X-band data are available, the automatization of the VT data analysis can also improve the rapid dissemination of crucial results to trigger the NIR follow-up; (ii) the NIR follow-up will benefit from new facilities, with the start of the SOXS spectrograph \citep{Soxs} operations in 2026, and the commissioning in spring 2026 of the CAGIRE camera \citep{Fortin2024} at the Colibrí telescope focus, that will observe all \textit{SVOM}\ GRBs visible from the ESO/La Silla and San Pedro M\'artir observatory, respectively. This is a considerable addition to the only handful of NIR instruments available to date for GRB follow-up. This optimized follow-up strategy will be the key to obtaining NIR spectroscopy at earlier times than for GRB\,250314A, so as to provide high-SNR spectra and fully exploit the possibility to probe high-$z$ galaxies and reionization with GRBs.

\begin{acknowledgements}
Acknowledgements can be found in the appendix.
\end{acknowledgements}

\vspace{-6ex}


\bibliographystyle{aa} 
\bibliography{main.bib} 

\newpage

J.~F.~Ag\"u\'i~Fern\'andez$^{21}$,
M.~A.~Aloy$^{22,23}$,
J.~An$^{2}$,
M.~Bai$^{24}$
S.~Basa$^{10}$,
M.~G.~Bernardini$^{25,26}$,
A.~Bochenek$^{27}$,
R.~Brivio$^{25}$,
M.~Brunet$^{12}$,
G.~Bruni$^{17}$,
S.~B.~Cenko$^{28}$,
Q.~Cheng$^{2}$,
A.~Chrimes$^{29,9}$,
L.~Christensen$^{7,8}$,
A.~Claret$^{1}$,
A.~Coleiro$^{30}$,
L.~Cotter$^{19}$, 
S.~Crepaldi$^{31}$,
J.~S.~Deng$^{2,3}$,
Dimple$^{32}$,
Y.~W.~Dong$^{15}$,
D.~Dornic$^{33}$,
P.~A.~E.~Evans$^{4}$,
R.~A.~J.~Eyles-Ferris$^{4}$,
H.~Fausey$^{34}$,      
M.~Ferro$^{25}$,
L.~Galbany$^{35,36}$,
M.~Garnichey$^{5}$,
G. Gianfagna$^{17}$,
B.~P.~Gompertz$^{32}$,
H.~Goto$^{37,11}$,
N.~Habeeb$^{4}$,
P.~Y.~Han$^{38}$,
X.~H.~Han$^{2}$,
D.~H.~Hartmann$^{39}$,
K.~E.~Heintz$^{7,8}$,
J.~Y.~Hu$^{2}$,
M.~H.~Huang$^{2}$,
L.~Izzo$^{40,41}$,     
P.~Jakobsson$^{42}$,
J.~A.~Kennea$^{43}$,
C.~Lachaud$^{30}$,
T.~Laskar$^{44}$,
D.~Li$^{45}$,
H.~L.~Li$^{2}$,
R.~Z.~Li$^{46}$,
X.~Liu$^{2}$, 
Y.~Liu$^{24}$,
G.~Lombardi$^{47,48,49}$,
H.~Louvin$^{1}$,
P.~Maggi$^{50}$,
T.~Maiolino$^{26}$,
Q.~Y.~Mao$^{45}$,
A.~Martin-Carrillo$^{19}$,
K.~Mercier$^{31}$,
P.~O'Brien$^{4}$,
J.~T.~Palmerio$^{11}$,  
P.~Petitjean$^{6}$,
D.~L.~A.~Pieterse$^{9}$,
F.~Piron$^{26}$,
G.~Pugliese$^{51}$,    
B.~C.~Rayson$^{4}$,
T.~Reynolds$^{7,8}$,
F.~Robinet$^{52}$
A.~Rossi$^{53}$,
R.~Salvaterra$^{54}$,
C.~C.~Th\"one$^{55}$,
B.~Top\c{c}u$^{56}$,
C.~W.~Wang$^{15}$,
J.~Wang$^{2}$,
Y.~Wang$^{57}$,
C.~Wu$^{2}$,
S.~L.~Xiong$^{15}$,
D.~Xu$^{2}$, 
H.~N.~Yang$^{2}$,
W.~M.~Yuan$^{2}$,
Y.~H.~Zhang$^{45}$,
X.~F.~Zhang$^{45}$,
S.~J.~Zheng$^{15}$

\vspace{2cm}

\small

$^{21}$Centro Astron\`omico Hispano en Andaluc\`ia, Observatorio de Calar Alto, Sierra de los Filabres, G\`ergal, Almer\`ia, 04550, Spain

$^{22}$Departament d’Astonomia i Astrof\`isica, Universitat de Val\`encia, C/Dr. Moliner, 50, E-46100 Burjassot (Val\`encia), Spain

$^{23}$Observatori Astron\`omic, Universitat de Val\`encia, 46980 Paterna, Spain

$^{24}$National Space Science Center, Chinese Academy of Sciences,Beijing 100190, China

$^{25}$INAF – Osservatorio Astronomico di Brera, Via Bianchi 46, I-23807 Merate, LC, Italy

$^{26}$Laboratoire Univers et Particules de Montpellier, Universit\'e Montpellier, CNRS/IN2P3, F-34095 Montpellier, France

$^{27}$Astrophysics Research Institute, Liverpool John Moores University, Liverpool Science Park, 146 Brownlow Hill, Liverpool L3 5RF, United Kingdom

$^{28}$Astrophysics Science Division, NASA Goddard Space Flight Center, Mail Code 661, Greenbelt, MD 20771, USA

$^{29}$European Space Agency (ESA), European Space Research and Technology Centre (ESTEC), Keplerlaan 1, 2201 AZ Noordwijk, The Netherlands

$^{30}$Universit\'e Paris Cit\'e, CNRS, Astroparticule et Cosmologie, F-75013 Paris, France

$^{31}$Centre National d’Etudes Spatiales, Centre Spatial de Toulouse, Toulouse Cedex 9, France

$^{32}$School of Physics and Astronomy and Institute for Gravitational Wave Astronomy, University of Birmingham, Birmingham B15 2TT, UK

$^{33}$Aix Marseille Univ, CNRS/IN2P3, CPPM, Marseille, France

$^{34}$Department of Physics and Astronomy, Baylor University, One Bear Place \#97316, Waco, TX, 76798, USA

$^{35}$Institut d’Estudis Espacials de Catalunya (IEEC), E-08034 Barcelona, Spain

$^{36}$Institute of Space Sciences (ICE, CSIC), Campus UAB, Carrer de Can Magrans, s/n, E-08193 Barcelona, Spain

$^{37}$College of Science and Engineering, School of Mathematics and Physics, Kanazawa University, Kakuma, Kanazawa, 9201192, Ishikawa, Japan

$^{38}$Department of Astronomy, School of Physics, Huazhong University of Science and Technology, 1037 Luoyu Road, Wuhan, 430074, People's Republic of China

$^{39}$Department of Physics \& Astronomy, Clemson University, Clemson, SC 29634, USA

$^{40}$INAF, Osservatorio Astronomico di Capodimonte, Salita Moiariello 16, I-80121 Naples, Italy

$^{41}$DARK, Niels Bohr Institute, University of Copenhagen, Jagtvej 155A, 2200 Copenhagen, Denmark

$^{42}$Centre for Astrophysics and Cosmology, Science Institute, University of Iceland, Dunhagi 5, 107 Reykjavik, Iceland

$^{43}$Department of Astronomy and Astrophysics, The Pennsylvania State University, 525 Davey Lab, University Park, PA 16802, USA

$^{44}$Department of Physics \& Astronomy, University of Utah, Salt Lake City, UT 84112, USA

$^{45}$Key Lab for Satellite Digitalization  Technology, Innovation Academy for Microsatellites, Chinese Academy of Sciences, No. 1 Xueyang Road, Shanghai 201304, China

$^{46}$Yunnan Observatories, Chinese Academy of Sciences, Kunming 650011, China

$^{47}$GRANTECAN S.A., Cuesta de San Jos\'e s/n, E-38712 Bre\~na Baja, La Palma, Spain

$^{48}$Instituto de Astrof\'isica de Canarias, E-38205 La Laguna, Tenerife, Spain

$^{49}$Royal Commission for AlUla, Al Safarat 12512, Riyadh, Saudi Arabia

$^{50}$Observatoire Astronomique de Strasbourg, Universit\'e de Strasbourg, CNRS, 11 rue de l’Universit\'e, F-67000 Strasbourg, France

$^{51}$Anton Pannekoek Institute of Astronomy, University of Amsterdam, P.O. Box 94249, 1090 GE Amsterdam, The Netherlands

$^{52}$Universit\'e Paris-Saclay, CNRS/IN2P3, IJCLab, 91405 Orsay, France

$^{53}$Osservatorio di Astrofisica e Scienza dello Spazio, INAF, Via Piero Gobetti 93/3, Bologna, 40129, Italy

$^{54}$INAF—Istituto di Astrofisica Spaziale e Fisica Cosmica di Milano, Via A. Corti 12, 20133 Milano, Italy

$^{55}$E. Kharadze Georgian National Astrophysical Observatory, Mt. Kanobili, Abastumani 0301, Adigeni, Georgia

$^{56}$Departmant of Physics, University of Bath, Claverton Down, Bath BA2 7AY, UK

$^{57}$Purple Mountain Observatory, Chinese Academy of Sciences, Nanjing 210023, China

\begin{appendix}

\section{Trigger, first localization and data transmission}

\subsection{Data transmission: VHF and X band data}
\label{appendix:vhf_xband}

\textit{SVOM} data are received on ground via two different channels: (i) selected data can be transmitted in near-real time via a VHF network (hereafter ``VHF data''); (ii) complete data are transmitted with a delay of a few hours, depending on the frequency with which X-band stations are available for the \textit{SVOM} mission (hereafter ``X-band data''). 
During the sequence following an on-board trigger, VHF data include alert packets (Appendix~\ref{appendix:trigger}) and selected data which are both generated by the on-board processing pipelines and used for the quicklook analysis on-ground.
In case of a slew of the spacecraft, VHF data include
an early photon list for the MXT, and source lists and 1-bit images for the VT, both around the ECLAIRs localization. X-band data include event-by-event data from ECLAIRs and GRM and all images obtained by MXT and VT, and are processed on-ground.
In the case of GRB\,250314A, MXT and VT VHF data were received 15~min and 49~min after the ECLAIRs trigger respectively, and X-band data were received with a delay of 4.7 hours.

\begin{figure}[h]
    \centering
    \includegraphics[width=1\linewidth]{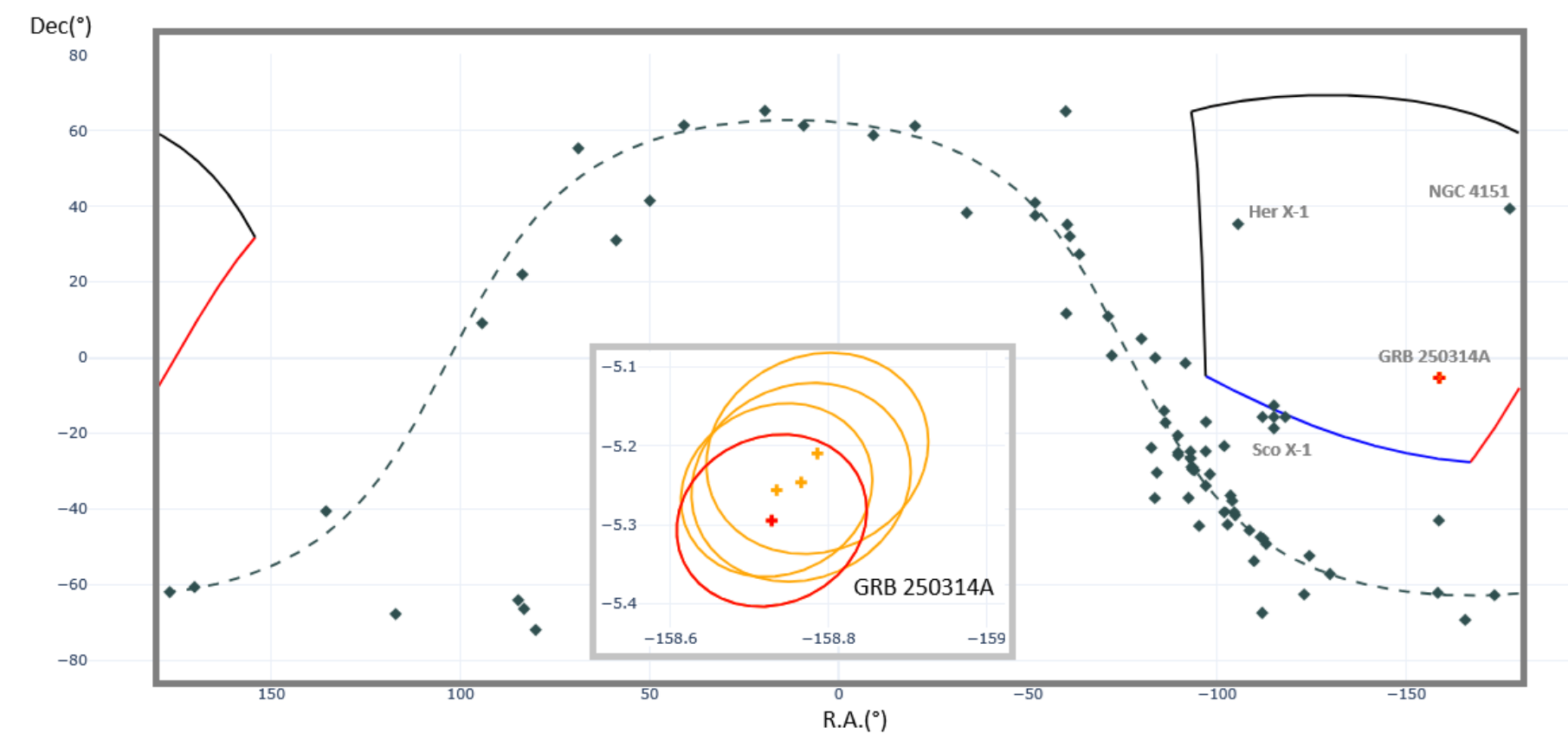}
    \caption{Location of the burst on the sky in Galactic coordinates (red point). The 4 alert packets produced by the trigger (inset) have coherent source positions, located well inside the 90 \% c.l. error ellipses.}
    \label{fig:trigger-alertpositions}
\end{figure}

\begin{figure}[h]
    \centering
    \includegraphics[width=1\linewidth]{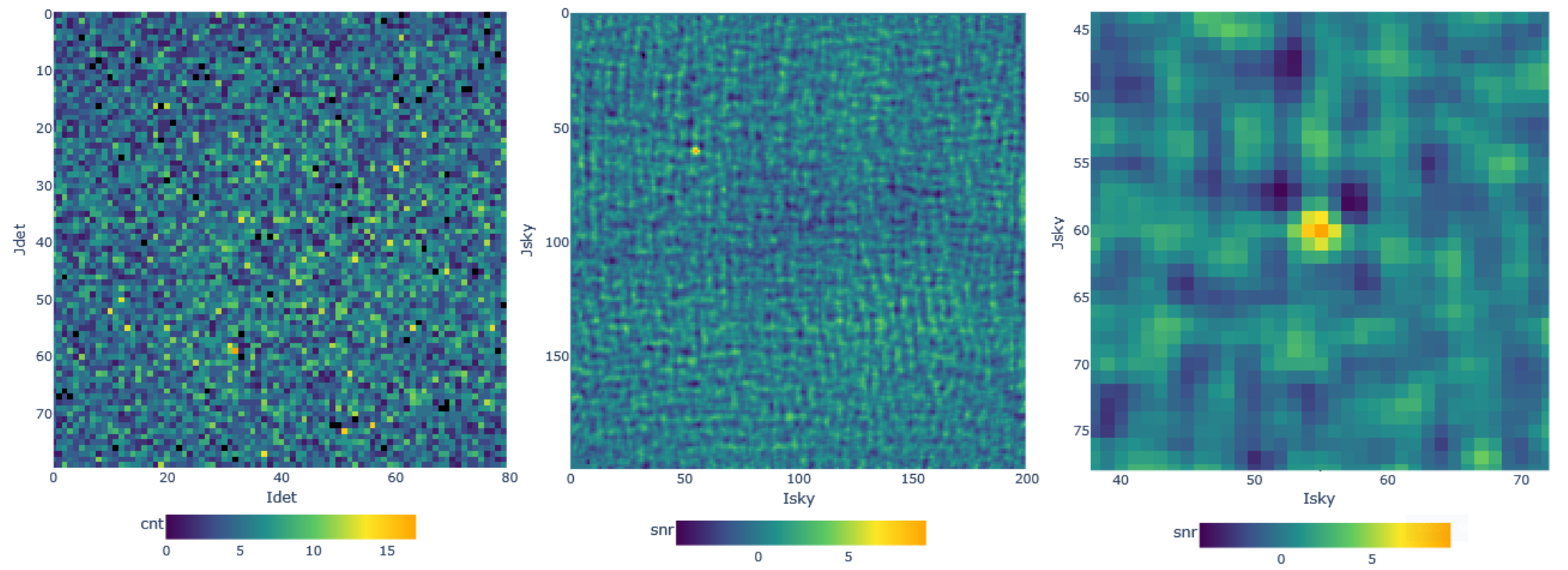}
    \caption{\textit{SVOM}/ECLAIRs VHF shadowgram (left) showing the number of counts per detector plane pixel during the time period of the best alert. Deconvolved full sky image (middle) in units of
    SNR ($\sigma$).
    Zoom on the detected source in the sky image (right).}
    \label{fig:trigger-shadowgram}
\end{figure}

\subsection{ECLAIRs on-board trigger and localization}
\label{appendix:trigger}

The \textit{SVOM}/ECLAIRs onboard trigger automatically detected and localized GRB\,250314A and produced 4 different VHF alerts packets, sent to the onboard VHF emitter.
The first VHF alert was produced at 12:56:58 UTC (T0) and covered the observation starting at 12:56:42 (Tb) over 10.24 s in the 8--50 keV energy band. 
The second VHF alert covered the same time range in the 8--120 keV band and provided the best localization, reported in Sect.~\ref{sec:obs}, with a maximum signal-to-noise-ratio (SNR) of 9.18 $\sigma$ (4.41 $\sigma$ above noise). The localizations of all 4 alerts were consistent (Fig. \ref{fig:trigger-alertpositions}). 
Those alerts were produced by the Count-Rate Trigger (CRT), an algorithm which searches for time periods with detector-count excesses over background, followed by the reconstruction of their sky image, in which a new and uncatalogued source is searched for \citep{Schanne2019}.
The trigger requested the spacecraft slew at T0+27~s and terminated the alert sequence 6~s later after confirmation of the slew start. It also sent VHF light-curve and VHF shadowgram packets, containing the detector plane image of the best alert (Fig.~\ref{fig:trigger-shadowgram}).

\section{ECLAIRs and GRM data analysis}
\label{appendix:prompt}

\subsection{GRM detection}
Since GRB~250314A triggered ECLAIRs first, GRM followed with an observation sequence driven by ECLAIRs. The analysis of the GRM X-band data 
shows a clear detection consistent with the ECLAIRs one, with high SNR in a timescale of 2\,s.

\subsection{Lightcurve and duration}
\label{appendix:prompt_T90}

Most of the burst emission in ECLAIRs is detected between 4 and 100 keV. The background-substracted lightcurve
in Fig.~\ref{fig:lightcurve} (top)
was built during the analysis of X-band data by selecting only pixels of the detection plane illuminated by the burst over a specific time window from 
Tb-109\,s to Tb+65\,s. 
This approach effectively minimizes background noise, thereby enhancing the accuracy of the background-subtracted light curves and facilitating a more precise estimation of the event duration. 
The background-subtracted lightcurve obtained from the analysis of GRM X-band data (Fig.~\ref{fig:lightcurve}, bottom) also consists of a single pulse. The T90 durations reported in Sect.~\ref{sec:obs} were determined by calculating the median of their distributions. These distributions were generated 
through simulations that employed Poisson resampling of the observed light curves.

\begin{figure}[h]
\centering
\includegraphics*[viewport = 0cm 0.25cm 15cm 11cm, width=0.85\linewidth]{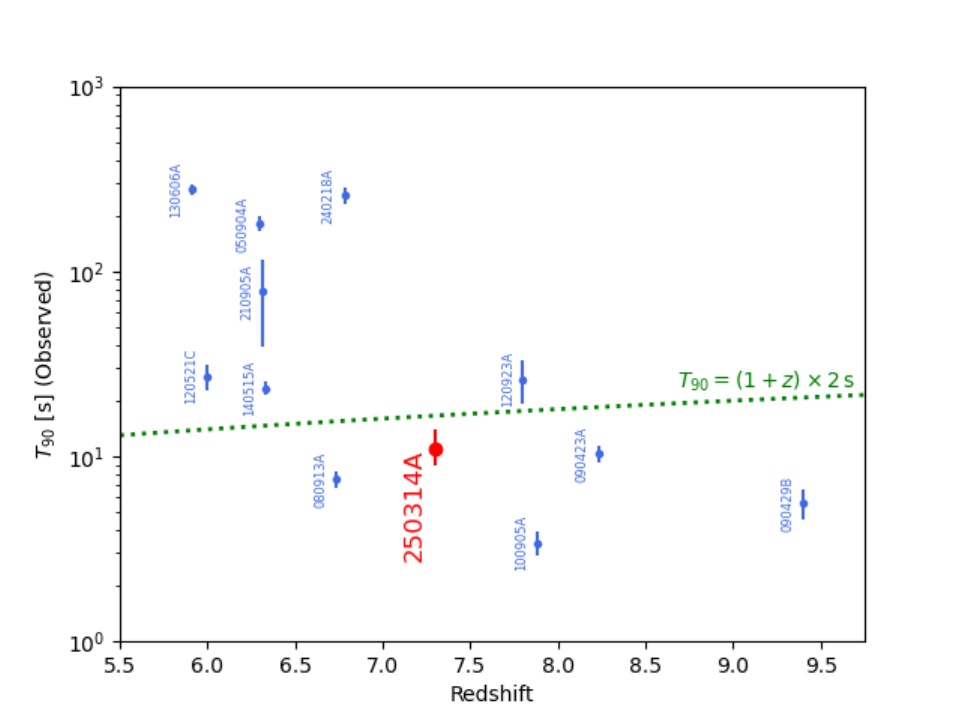}
\caption{Observed T90 duration as a function of the redshift for the current sample of GRBs above $z\sim 6$. The observed T90 duration in 4--100 keV of GRB\,250314A is shown in red. The observed T90 durations in 15--150 keV of the other bursts in blue are taken from the {\it Swift}/BAT catalog \citep{lien:16}. The dotted green line corresponds to a duration of 2 s in the burst rest frame.}
\label{fig:duration}
\end{figure}

The result $T_{90}=11^{+3}_{-2}$ s in the 4--100 keV band of ECLAIRs
corresponds to a rather short T90 duration in the burst rest frame ($1.3^{+0.4}_{-0.2}\, \mathrm{s}$). The measured duration is however very sensitive to the energy range and sensitivity of the detector, and the burst SNR (see e.g. the discussion for \textit{Swift}/BAT in \citealt{moss:22}). 
Such short durations are observed in most GRBs above $z=7$ as shown in Fig.~\ref{fig:duration} and can be interpreted as a ``tip of the iceberg effect'', as discussed in \citet{littlejohns:13,lu:14}. Such an effect is also expected for high redshift GRBs detected by ECLAIRs as shown in the pre-launch simulations performed by  \citet{llamas24} (see their Fig.~3).

\subsection{Spectral analysis}
\label{appendix:prompt_spectral}

The best time interval to build the  spectrum of GRB~250314A
was found using reconstructed ECLAIRs sky images over different time intervals so that the source was detected with a signal-to-noise ratio larger than~3.
This leads to an interval of 11~s from Tb-1 s to Tb+10 s, used  for the analysis of the time-averaged spectrum by ECLAIRs, GRM and ECLAIRs+GRM. 

\textit{ECLAIRs spectral analysis.} We extracted the time-averaged spectrum
using the ECLAIRs pipeline (ECPI) version 1.18.0. ECPI first produces Good Time Intervals (GTIs) by filtering auxiliary and housekeeping data (i.e. filtering out satellite passages through the SAA, excluding time intervals when the satellite pointing is unstable, and selecting data according to Earth occultation conditions). The pipeline discards all events exhibiting abnormal Pulse Height Analyser (PHA) values, each event being identified by its pixel position and associated energy channel. Then, the event energy is calculated using a linear calibration function that accounts for the pixel-specific gain and offset parameters and  is assigned to a unique Pulse Invariant (PI) channel. ECPI corrects detector images produced in different energy bands for efficiency, non-uniformity, and Earth occultation. From these corrected detector images, the pipeline applies a deconvolution procedure to reconstruct sky images and performs coding noise cleaning. 

In addition, based to the burst position in these sky images, the pipeline builds a specific shadowgram model. The shadowgram model is fitted to the detector maps to extract the flux information for each energy bin. A second-degree polynomial is used to account for the background component. The response matrix file ECL-RSP-RMF\_20220515T103600.fits and the ancillary response file ECL-RSP-ARF\_20220515T104100.fits were used in the analysis.
The ECLAIRs time-averaged spectrum is best fitted ($\chi^{2}/\mathrm{d.o.f.} = 19/21$) in the 5--100 keV band by a powerlaw with a photon index of $1.2 \pm 0.1$ 
corresponding to a 4–120 keV time-averaged flux of 5.1$^{+0.5}_{-0.9}\times 10^{-8}{\,\rm erg\,cm^{-2}\,s^{-1}}$.

\textit{GRM spectral analysis.} We used all three Gamma-Ray Detectors (GRDs) to extract the time-averaged spectrum.
The Earth atmosphere albedo is not taken into account in the responses as
the orientation of the three GRDs are all away from the Earth, and the flux of the burst is relatively low.
The GRM time-averaged spectrum is best-fitted
by a cutoff powerlaw model
(CPL), with parameters of $\alpha=1.00^{+0.57}_{-0.67}$ and $E_p=64^{+16}_{-14}$\,keV
(stat/d.o.f.$\sim$34/24).
This corresponds to a time-averaged flux of 3.27$^{+0.56}_{-0.55}\times10^{-8}{\,\rm erg\,cm^{-2}\,s^{-1}}$ 
in the 15-5000 keV band.

\begin{figure}[b]
    \centering
    \includegraphics[width=0.8\linewidth]{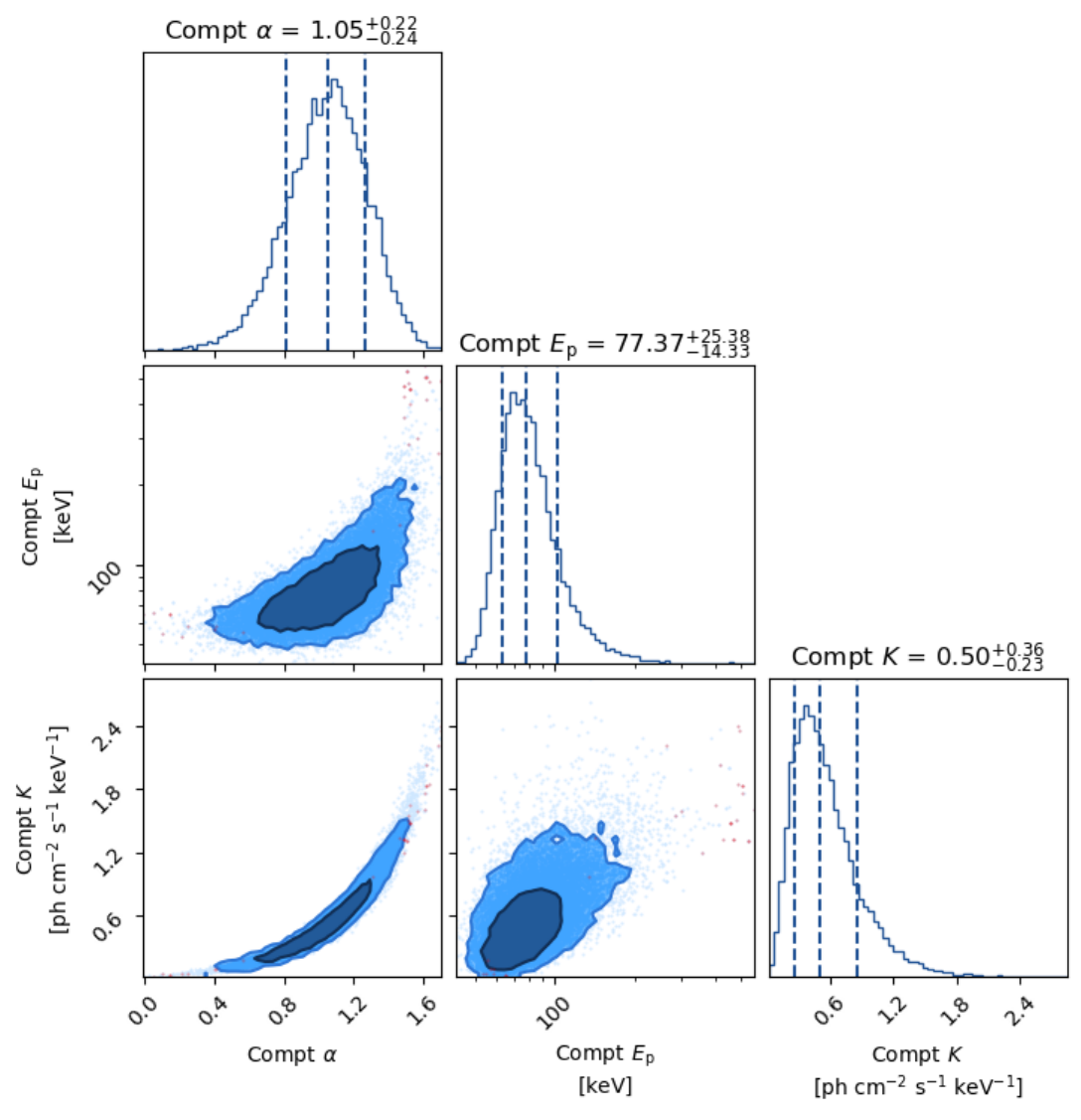}
    \caption{Best-fit parameters of the ECLAIRs and GRM joint spectral fits using a CPL model in the 5 keV -- 5 MeV energy band. Errors are quoted at the $1\,\sigma$ level.}
    \label{fig:spectral_parameters}
\end{figure}

\textit{ECLAIRs+GRM joint spectral analysis.} We then performed a joint spectral analysis in the 5 keV--5 MeV energy band. The time-averaged spectrum is best fitted by a CPL model with a photon index $\alpha = -1.05^{+0.22}_{-0.24}$, 
a peak energy ${E}_\mathrm{p}=77^{+25}_{-14}$ keV 
and a normalization $\kappa = 0.50^{+0.36}_{-0.23}$ ph cm$^{-2}$ s$^{-1}$ keV$^{-1}$ at 1 keV (stat/d.o.f. = 152/147).
The posterior distributions of the model parameters are shown in Fig.~\ref{fig:spectral_parameters}.
Note that given the burst weakness 
the peak energy
is not well constrained.

\section{X-ray data analysis}
\label{appendix:xrays}

\subsection{\textit{SVOM}/MXT data analysis}

 MXT data were acquired in the standard event mode and processed with the MXT pipeline v1.12. The MXT on board software, attempting localizations on increasingly longer exposure times, starting from 30 s after satellite stabilization, did not find any new source in the MXT field of view in near-real time. As MXT did not detect the afterglow of GRB\,250314A
 we determined the two  3 $\sigma$ upper limits on the 0.3--10 keV energy band reported in Tab.~\ref{tab:obs_Xray}. The first covers the first ks of observation, and the second corresponds to the entire observation. Both have been calculated assuming the spectral parameters measured by \textit{Swift}/XRT and \textit{EP}/FXT, see Appendix~\ref{appendix:xrt_fxt}.

\begin{figure}[h]
    \centering

    \includegraphics[scale=0.5]{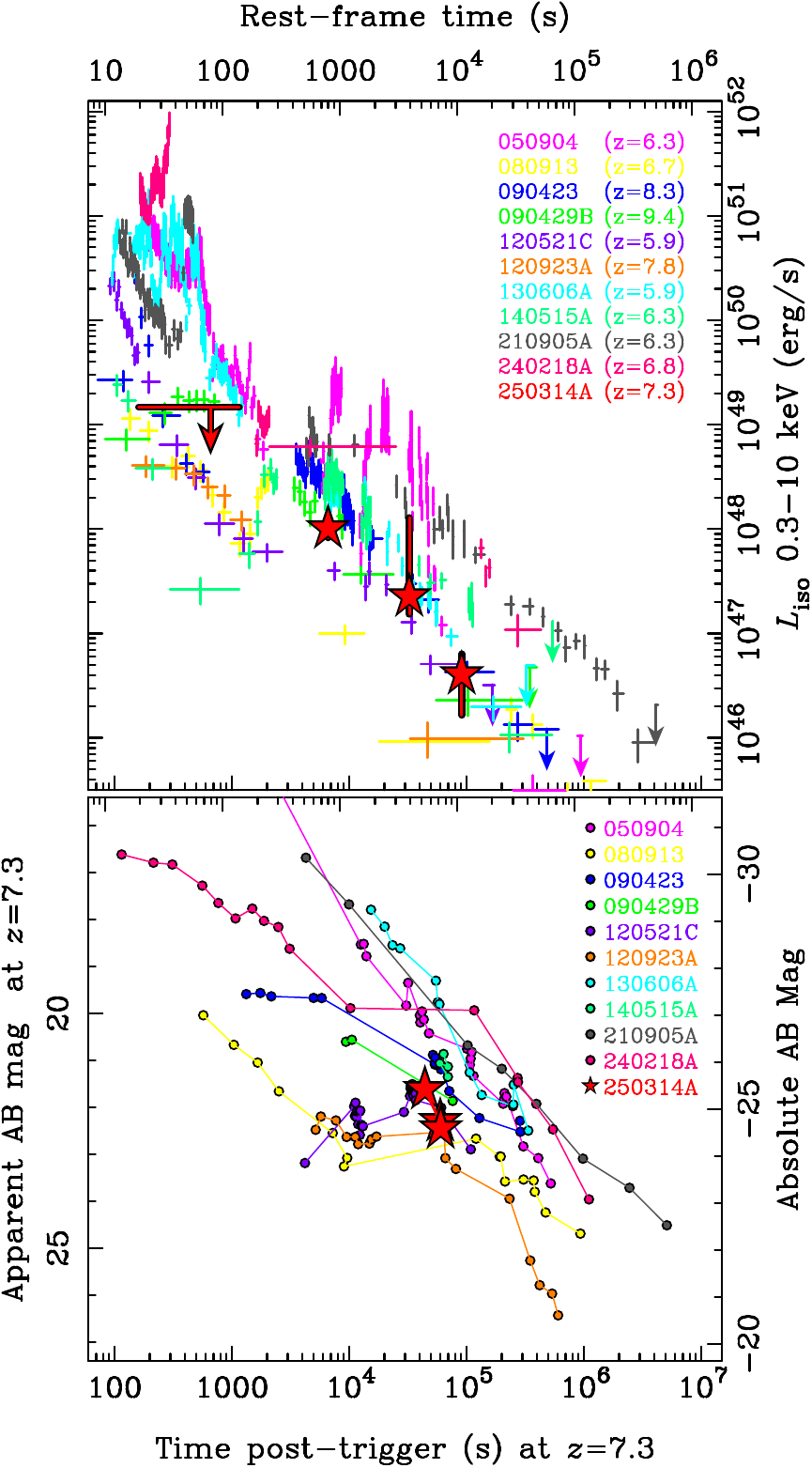}
    \caption{Comparison of the afterglow lightcurves of a sample of high-z GRBs (color coded) to our data (cyan stars).
    \textit{Top:} unabsorbed X-ray luminosity (0.3-10 keV), deduced for GRB250314A from the \textit{SVOM}/MXT, \textit{Swift}/XRT and \textit{Einstein Probe}/FXT data in Tab.~\ref{tab:obs_Xray}, and obtained through the {\it Swift} burst analyzer \citep{Evans_10}
    for the other high-z GRBs.
     \textit{Bottom:} Optical lightcurves. Each rest-frame light curve is shifted to a common $z=7.3$ redshift and to H-band. Only detections redward of the Ly-$\alpha$ break have been included. Optical light curves of high-z afterglows are taken from \citet{Tanvir2018} and references therein, \citet{rossi:22,brivio:25}.
    }
    \label{fig:afterglow_lightcurve}
\end{figure}

\subsection{\textit{Swift}/XRT and \textit{EP}/FXT data analysis}
\label{appendix:xrt_fxt}

\begin{table}[b]
\tiny
\caption{X-ray follow-up observations. 
The flux are expressed in the 0.3--10 keV energy band and upper limits are given at the $3\sigma$ level.}
\label{tab:obs_Xray}
\vspace*{-3ex}

\begin{center}
\resizebox{\linewidth}{!}{
\begin{tabular}{c|cccc}
\hline \hline
Instrument & Time interval & \hspace*{-0.5cm} Exposure\hspace*{-0.5cm}  & \multicolumn{2}{Sc}{Flux}   \\
& since Tb &  time    & Observed & Unabsorbed   \\
 	           & (s)	    & (ks)  & \multicolumn{2}{Sc}{($\mathrm{10^{-12}\cdot erg~cm^{-2}~s^{-1}}$)} \\
\hline
\addlinespace[1ex]
\multirow{2}{*}{\textit{SVOM}/MXT}
    &   177-1177 
    & 1.0  
    & \multicolumn{1}{Sc}{$\le25$}  \\        
    &   177-11966 
    &  4.95 
    & \multicolumn{1}{Sc}{$\le11$} \\
\addlinespace[1ex]
\hline
\addlinespace[1ex]
\multirow{2}{*}{\textit{Swift}/XRT}  
    & $\left(5.68-7.38\right)\times 10^3$ 
    & 1.69 
    &  $1.1^{+0.8}_{-1.0}$ 
    & $1.7^{+0.3}_{-0.3}$ \\
    & $\left(343.22 - 811.52\right)\times 10^{3}$ 
    & 3.91 
    &  \multicolumn{1}{Sc}{$\le 0.2$} \\  
\addlinespace[1ex]
\hline
\addlinespace[1ex]
\multirow{3}{*}{\textit{EP}/FXT}  
    &  $\left(31.97 - 33.16\right)\times 10^{3}$  
    & 1.19 
    &  $0.4^{+0.1}_{-0.1}$ 
    & $0.4^{+0.1}_{-0.1}$ \\
    \addlinespace[0.5ex]
    &  $\left(83.83 - 98.39\right)\times 10^{3}$ 
    & 8.87 & $0.06^{+0.04}_{-0.03}$  
    & $0.07^{+0.04}_{-0.03}$  \\
    &  $\left(250.93 - 259.73\right)\times 10^{3}$
    & 5.48  
    & \multicolumn{1}{Sc}{$\le 0.05$} \\
\addlinespace[1ex]
\hline
\hline
\end{tabular}
}
\end{center}
\end{table}

A first automatic \textit{Swift} ToO was executed at 14:31:21 UT, i.e. 1.6 hr after Tb, for a total exposure time of 1.7 ks. Additional observations were also carried out from 4.0 to 8.9 days after Tb.
Three epochs were obtained with \textit{EP}/FXT from 2025-03-14 21:49:29 to 2025-03-17 13:05:32 UT, i.e.
8.9\,h, 23.3\,h and 2.9\,days after Tb. 
We clearly detected an uncatalogued fading X-ray source in the first \textit{Swift}/XRT \citep{GCN_Swift} and \textit{EP}/FXT \citep{GCN_EP} observations at an enhanced \textit{Swift}/XRT position $\mbox{R.A.} = 13h 25m 12.33s$ and $\mbox{Dec} = -05^\circ 16^\prime56.1^{\prime\prime} $ (J2000) with an uncertainty of 3.4$^\prime{^\prime}$. This position is 1.9$^\prime$ away from the ECLAIRs position, and is identified as the X-ray afterglow of GRB~250314A due to its clear fading signature.

The \textit{Swift}/XRT spectra were obtained from the UK \textit{Swift} Science Data Centre\footnote{\url{https://www.swift.ac.uk/xrt_spectra/00019616}}, while the \textit{EP}/FXT (A and B) spectra were processed following the standard data reduction procedure described in the FXT Data Analysis Software Package\footnote{\url{http://epfxt.ihep.ac.cn/analysis}} (FXTDAS v1.10). Both count rate spectra were grouped to a minimum of 1 count per bin using the \textit{grppha} ftools routine. Data have been fitted using the XSPEC package\footnote{\url{https://heasarc.gsfc.nasa.gov/docs/xanadu/xspec/}}. We applied the \textit{C-statistic} method \citep{Cash1979} to fit our data. 
We first separately analysed the \textit{Swift}/XRT and \textit{EP}/FXT spectra of their first respective epochs. Due to the low statistics, (17 and 31 counts in FXT A and B, and 44 counts in XRT), we then attempted a joint fit by allowing for a constant between XRT and FXTs to compensate for the flux variability, and cross calibration uncertainties. 
In Tab.~\ref{tab:spec_Xray}, we report the results of our fits using an absorbed power law model ($Tbabs\times powerlaw$)\footnote{The abundances used are from \citet{anders89}} and using an additional local absorber ($Tbabs\times zTbabs \times powerlaw$) at $z=7.3$ (fixing the galactic hydrogen column density fixed at $N^{Gal}_{H,x} = 2.46\times 10^{20}cm^{-2}$ \citep{HI4PI2016}). The derived fluxes can be found in Tab.~\ref{tab:obs_Xray}. The top panel in
Fig.~\ref{fig:afterglow_lightcurve} shows the X-ray light curve of GRB~250314A compared to a sample of 12 high-$z$ X-ray afterglows.

In all cases the best fit is obtained by an absorbed power law, with $N_{H}$ fitted values in excess of the Galactic one. 
Interestingly, we notice that when trying to account for this excess by modelling it as the intrinsic host galaxy hydrogen column density contribution, we obtain values that are in excess of N$^{host}_{H,x}$ > 10$^{23}$ cm$^{-2}$. Such a large amount of $N^{host}_{H,x}$ has been previously reported for other high-$z$ GRBs, such as GRB\,090429B \citep{Cucchiara2011} and GRB\,090423 \citep{Salvaterra2009}, and could be due to an additional contribution of the Intergalactic Medium (IGM), 
as pointed out by \citet{starling13}. 
However the low statistics hinders further speculation on this point.

\begin{table}[h]
\centering
\tiny
\caption{Results of the spectral analysis of the \textit{Swift}/XRT and \textit{Einstein Probe}/FXT X-ray follow-up observations. 
$N^{\rm Gal}_{H,X}$ is extracted from the HI4PI Map \citep{HI4PI2016} at the R.A., Dec. coordinates of the XRT afterglow.}
\label{tab:spec_Xray}
\begin{tabular}{c|ccccc}
\hline \hline

Instrument	  & Time 
& $N_{H,X}$     & $N^{\rm host}_{H,X}{\dagger}$     & Photon  & C-stat/{d.o.f.} \\[1ex]
          & since Tb	    &  & & index   &   \\
 	   & (ks)	    
        & \multicolumn{2}{c}{($10^{22}\, \mathrm{cm^{-2}}$)} 

       & ($\Gamma$)   & (d.o.f.)  \\

\hline
\addlinespace[1ex]
{\textit{Swift}/XRT}  
    & {5.68} 
    & 0.19$^{+0.22}_{-0.17}$& - & 2.9$^{+1.2}_{-0.9}$ & 0.99 (39) \\
    &  7.38 
    &    0.0246   & 34$^{+49}_{-26}$ & $2.1\pm0.3$ & 0.91 (39) \\[1ex]
                \hline
\addlinespace[1ex]

{\textit{EP}/FXT}  
    &  31.97 
    & 0.10$^{+0.39}_{-0.10}$ & - & 2.4$^{+2.2}_{-1.1}$ & 1.41 (44) \\
(A+B)        
    &  33.16 
    &    0.0246  & 4$^{+22}_{-4}$ & 2.1$^{+1.1}_{-0.7}$  & 1.41 (44) \\[1ex]
   \hline
\addlinespace[1ex]
{XRT/FXT}  
    &  {5.68} 
    & $<$0.35 & - & 1.8$^{+2.0}_{-0.4}$  & 0.51 (78) \\
  (A+B) joint        
    &  33.16 
    &     0.0246   & 21$^{+32}_{-18}$ & 2.7$^{+0.9}_{-0.7}$  & 0.9 (77) \\[1ex]
\hline
\hline
\end{tabular}
\begin{tablenotes}

      \item $^\dagger$: With additional $N^{Gal}_{H,x}$ fixed at $2.46\times 10^{20}$cm$^{-2}$ and $z=7.3$.
    \end{tablenotes}
\end{table}

\section{VT data reduction}
\label{appendix:vt}
VT follow-up observations lasted for 3 orbits continuously during the full Moon phase. Although the angular distance to the Moon was 25$^\circ$, the stray light strength from the Moon in VT images was negligible.

\subsection{Quicklook analysis}

The on-board data processing pipeline processed the images with four sequences,
two performed during the first orbit,
and two during the second orbit, aiming to generate the quick look products to have a rapid identification of an optical counterpart. 
During the processing, the quality of images were checked mainly based on the stability of the platform and background level. The instrument calibrations including bias, dark and LED flat-field correction were then carried out.

Using subimages centered at the ECLAIRs localization with a size covering partially the ECLAIRs errorbox, the on-board processing pipeline generates source lists for the four sequences and 1-bit images for the first and second sequences (MXT localization would have been used if it had been available; in this case the subimages size cover the entire MXT errorbox).
The source lists contain the sources above the detection threshold, which is configurable on-board with a default value of 2$\sigma$ (SNR per pixel).
The 1-bit images are combined subimages digitized with 0 or 1 (1-bit) depending if the pixel is below or above the detection threshold.
This is adapted to the limited data rate of the VHF network.
After being encoded with the Run-Length Encoding method, 1-bit images are near two orders of magnitude smaller than the original images (16 bits).
In the case of GRB\,250314A, the subimage transmitted to the ground by the VHF network
was generated by stacking six consecutive images
and was covering $\sim$ 55\% of the ECLAIRs error-box.

Once the VHF data (source list and 1-bit subimage) were received, the on-ground real-time data processing pipeline made the astrometric and photometric calibrations. The source lists were cross-checked with known catalogued sources, with a visual check by scientists on duty.
The results of this quicklook analysis suggested with high confidence level that there are no uncatalogued sources brighter than 20 mag \citep{GCN_VT_VHF}.

\subsection{X-band data analysis: deep optical upper limits.}

\begin{figure}[h]
    \centering
    \includegraphics[scale=0.2]{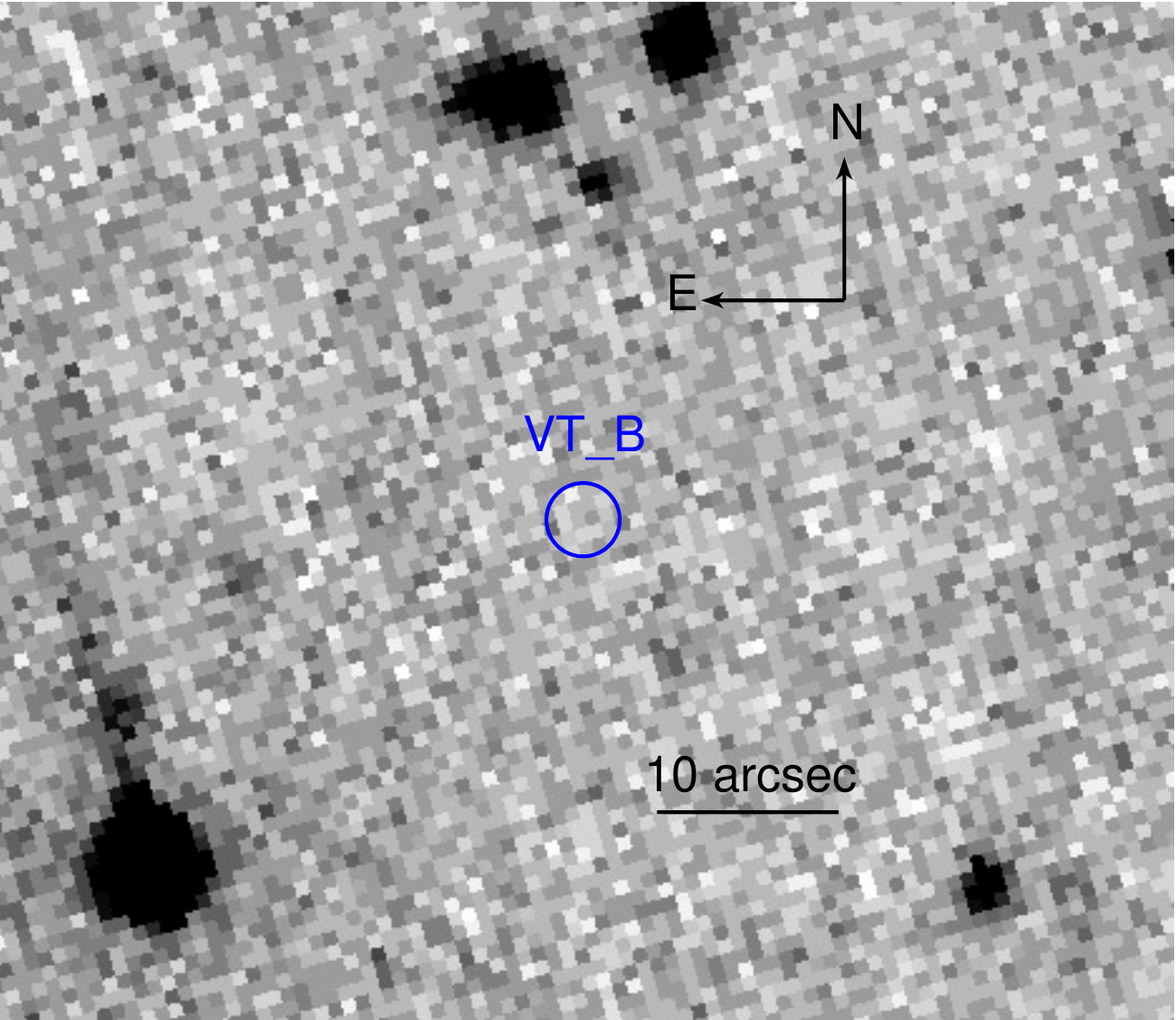}
    \includegraphics[scale=0.2]{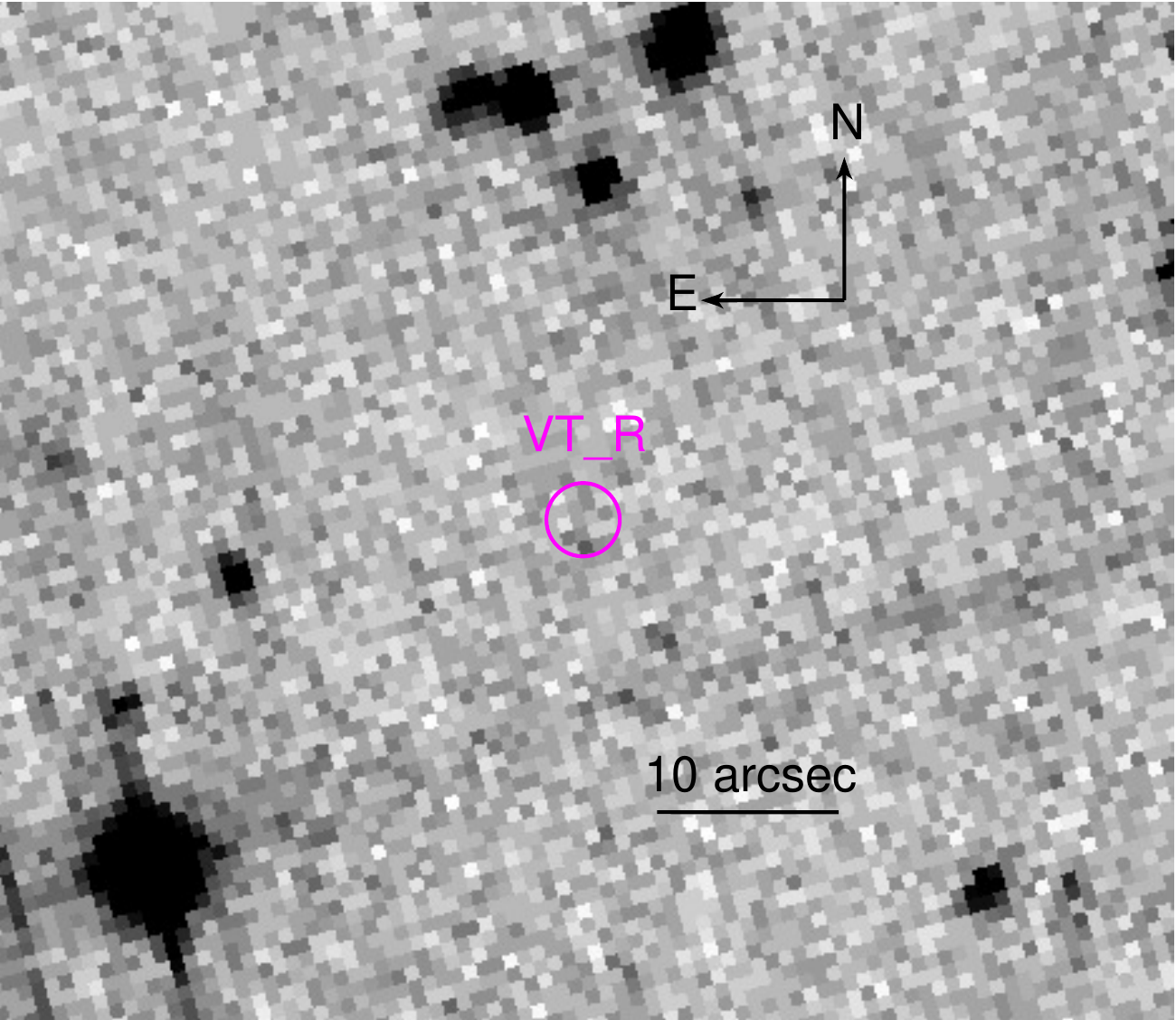}
    \caption{VT stacked image in VT$\_$B and VT$\_$R bands. The total effective exposure time is 24$\times$50 s and 23$\times$50 s for VT$\_$B and VT$\_$R bands, respectively, at a mid time of 13.9 min after the burst. The upper is north and the left is the east. the central red circle is the location of GRB 250314A. No signal was detected. }
    \label{fig:vt-img}
\end{figure}

When X-band data were  received later, a more refined analysis was performed, by stacking more calibrated images.
Focusing on the precise localization provided by \textit{Swift}/XRT \citep{GCN_Swift}, and after excluding  fake sources caused by detector defects and faint catalogued sources, no optical candidate was detected down to the deep 3 sigma upper limits provided in Tab.~\ref{VT_upperlimit}. 
The corresponding finding chart for the stacked image
in VT\_R channel (650-1000~nm)
is shown in Fig.~\ref{fig:vt-img}.

\begin{table}[h]  
\small
\centering
\caption{Log of VT 3$\sigma$ upper limits for GRB\,250314A. 
} 
\begin{tabular}{c c c c c}       
\hline
Time since Tb
& Exptime (s) & Upper limit (AB) &  Band \\
\hline
13.9 min & 24$\times$50  & 23.5 & VT\_B \\
13.9 min & 23$\times$50  & 23.0 & VT\_R \\
1.0 hr & 79$\times$50  & 23.8 & VT\_B \\
1.0 hr & 77$\times$50  & 23.2 & VT\_R \\
\hline
\end{tabular}
\label{VT_upperlimit} 
\end{table}

\section{Ground-based optical-NIR follow-up}
\label{appendix:v_nir}

\subsection{Photometry}

The data reduction of the VLT/HAWK-I observations reported in Sect.~\ref{sec:obs}
was carried out using the standard ESO pipeline. Near-IR photometric calibration was performed against the VHS catalog. The small lever arm between $J$ and $H$ does not provide strong constraints on the spectral slope, but formally we find $\beta=0.19\pm0.40$, where flux density is expressed as a power-law function of frequency, $F_{\nu}\propto\nu^{-\beta}$. This is relatively blue for a GRB afterglow, suggestive of a low extinction line of sight.

The much steeper slope between the $Y$ and $J$ bands, and deep non-detection in $z$ indicates a sharp spectral break around $\lambda\approx1\,\mu$m.

\subsection{Optical-NIR spectral analysis}

The VLT/X-shooter observations,
also reported in Sect.~\ref{sec:obs},
consisted of 4 exposures of 1200~s each. These observations were conducted using the ``ABBA'' nod-on-slit mode. Each individual VIS spectrum was reduced using the STARE mode reduction, while the NIR arm data was processed using the standard X-shooter NOD mode pipeline \citep{Modigliani2010}. Sky features were subtracted and each flux-calibrated spectrum combined into a final science product \citep{Selsing2019}.
All magnitudes are reported in Tab.~\ref{tab_photometry}.

To reveal the faint continuum
in the X-shooter spectrum  shown in Fig.~\ref{fig:VLT_spectrum} (left), we first masked bad pixels and subtraction residuals of bright sky lines, then binned the VIS (NIR) into 2.5 (1.5) nm wide channels using inverse variance weighting. The extraction was performed using weighting derived from a Gaussian fit to the spatial profile of the trace in the NIR arm. No correction was made for slit losses, although this correction is expected to be small, given the good seeing conditions ($\sim0.55$ arcsec).
As discussed in Sect.~\ref{sec:obs}, this analysis revealed a flat continuum with a sharp drop off consistent with the Lyman-$\alpha$ break due to absorption from neutral hydrogen, leading to the redshift determination, $z\simeq 7.3$. 
More careful fitting of the spectrum with a power-law continuum and Lyman-$\alpha$ damping wing, adopting a flat prior on the logarithm of the neutral hydrogen column of the host galaxy (i.e. uniform between $16<{\rm log}(N_{\rm H}/{\rm cm}^{-2})<25$) produces a refined redshift estimate of $z\approx7.27^{+0.02}_{-0.03}$.
The same modelling finds an upper bound to the host neutral hydrogen column of ${\rm log}(N_{\rm H}/{\rm cm}^{-2})<22.5$, but is unable to place a constraint on the neutral fraction of the intergalactic medium.

The spectral slope from the X-shooter NIR arm is measured as $\beta=0.42\pm0.14$, consistent with the blue slope found from the photometry.

Due to the very low signal-to-noise, no individual metal absorption features have been confidently identified. EW upper limits could be calculated in different regions of the NIR spectrum (with low atmospheric absorption and good transparency). The SNR is about $\sim0.4$ ($\sim0.3$) which translates into an EW upper limit rest-frame $<1.5$~\AA\, ($<2.6$~\AA), assuming a 3-$\sigma$ limit and a nominal X-shooter resolution in the NIR arm of 5600 at $12000$~\AA. These EW limits are above those typically found in GRB afterglow spectra for the host metal absorption lines detectable in this wavelength range \citep{deUgartePostigo2012}.

\begin{table}[h]     
\small
\centering
\caption{Optical and NIR ground-based photometry. $z$-band photometry was calibrated against the Pan-STARRS catalogue, and $Y$, $J$, and $H$ against the VISTA Hemisphere Survey \citep{mcmahon21}. Upper limits are given at the $3\sigma$ levels. No correction has been made for the small Galactic foreground extinction \citep[$A_{\rm J}=0.026$, $A_{\rm H}=0.016$;][]{Schlafly2011}.} 
\begin{tabular}{c c c c c}       
\hline\hline
Time & Exposure  & Instrument & Magnitude & Band\\
since Tb (hr) & time (s) & & (AB) & \\
\hline
16.7  & 360  & VLT/X-shooter & $>$23.1            & $z$ \\
16.81  & 930  & GTC/OSIRIS+   & $>$23.9            & $z$ \\
16.96  & 600  & VLT/HAWK-I    & 23.29$\pm$0.13     & $Y$\\
12.31  & 1080 & NOT/NOTCam    & 21.82$\pm$0.17     & $J$ \\
16.61  & 480  & VLT/HAWK-I    & 22.50$\pm$0.08     & $J$\\
38.64 & 2700 & NOT/NOTCam    & $>$21.5            & $J$ \\
16.74  & 480  & VLT/HAWK-I    & 22.44$\pm$0.06     & $H$\\
\hline
\hline
\end{tabular}
\label{tab_photometry} 
\end{table}

\section{Radio follow-up}
\label{appendix:radio}

VLA carried out a first set of  observations  7 days after Tb, leading to detections at 10 GHz and 15 GHz. The follow-up campaign in radio with ATCA, e-MERLIN and MeerKAT gathered then 10 observations from 9  to 109~days after Tb at several frequencies from 1.3 to 9 GHz.

\textit{ATCA data.} Following the determination of the spectroscopic redshift, we triggered radio observations with the Australian Telescope Compact Array under program C3546 (PI Thakur). Observations were obtained in the C (5.5 GHz) and X (9 GHz) bands at one epoch 8.96 days post-burst, for a total of 9.5 hours. The observations in the last two hours were strongly affected by technical issues, resulting in an increased noise level. The datasets were processed in CASA \citep{McMullin:07} using standard flagging, calibration and imaging procedures. The primary and bandpass calibrator was 1934-638, and the phase calibrator was 1308-098. Flux densities were evaluated at the source peak in the restored images, as this is the best estimate for point-like sources. Root Mean Square (RMS) noise in the final radio maps was computed in an annulus around the source position. The flux error was calculated as the squared sum of the image RMS plus a 5\% uncertainty on the flux scale calibration. The source was detected only at 9 GHz at the near-infrared position. It is not detected at 5.5 GHz, the corresponding 3-sigma upper limit is included in Tab.~\ref{tab:radio}. 

\textit{e-MERLIN data.}
We also requested Director’s Discretionary Time observations with the Enhanced Multi-Element Remotely Linked Interferometer Network (e-MERLIN) under program RR19004 (PI Thakur). They were carried out at C-band (5.1 GHz) and in two epochs, 16.46 and 56.46 days post-burst, including the following antennas: Mk2, Pi, Kn, De, Cm. The phase calibrator was 1332+0509, while 3C286 was adopted for amplitude calibration. The total durations of the two observing runs were $\sim36$ hours and $\sim24$ hours, respectively. Data was processed with the e-MERLIN pipeline \citep{Moldon:21}, and imaging was performed with CASA \citep{McMullin:07} at the central frequency of 5.1 GHz, adopting natural weighting. The RMS was measured in an annulus surrounding the target. The source is not detected at either epoch, the corresponding 3-sigma upper limits are given in Tab.~\ref{tab:radio}. 

\textit{MeerKAT data.}
We also triggered observations under the MeerKAT program SCI-20241102-AT-01 (PI Thakur) at L (1.3 GHz) and S2 (2.6 GHz) bands. Observations were conducted at 9.29, 29.2 and 109.10 days post burst, with 30 minutes on-source integration. Phase referencing using 1334-127 as calibrator was applied, while the flux scale calibrator was 1934-738. Data was reduced with the SARAO Science Data Processor (SDP) continuum pipeline. The resulting image has an angular resolution of 8 x 7 arcsec at L band, and 5 x 4 arcsec at S2 band. The average RMS noise level was 10 uJy/beam at L band and 4 uJy/beam at S2 band. The source was not detected at either band in any epoch, the corresponding 3-sigma upper limits are reported in Tab.~\ref{tab:radio}. 

\textit{VLA data.}
We observed GRB\,250314A on 2025 March 21.27 UT at a mean time of 6.7 days after the \textit{SVOM} trigger with the Karl G. Jansky Very Large Array (VLA) with project VLA/24B-072 (PI: Laskar) using the 3-bit receivers at C, X, and Ku bands (providing 4 GHz, 4 GHz, and 6 GHz bandwidth, respectively). The observations utilized 3C286 as the flux density and bandpass calibrator and J1337-1257 as complex gain calibrator. We reduced the data using the CASA VLA pipeline. Upon imaging the data, we detect a radio counterpart at Ku-band (center frequency, 15 GHz) and X-band (10\,GHz), but not at C-band (6 GHz). We report our point-source fitting photometry results and $3\sigma$ upper limits in Tab.~\ref{tab:radio}.

\begin{figure}[t]
    \centering
    \includegraphics[scale=0.43]{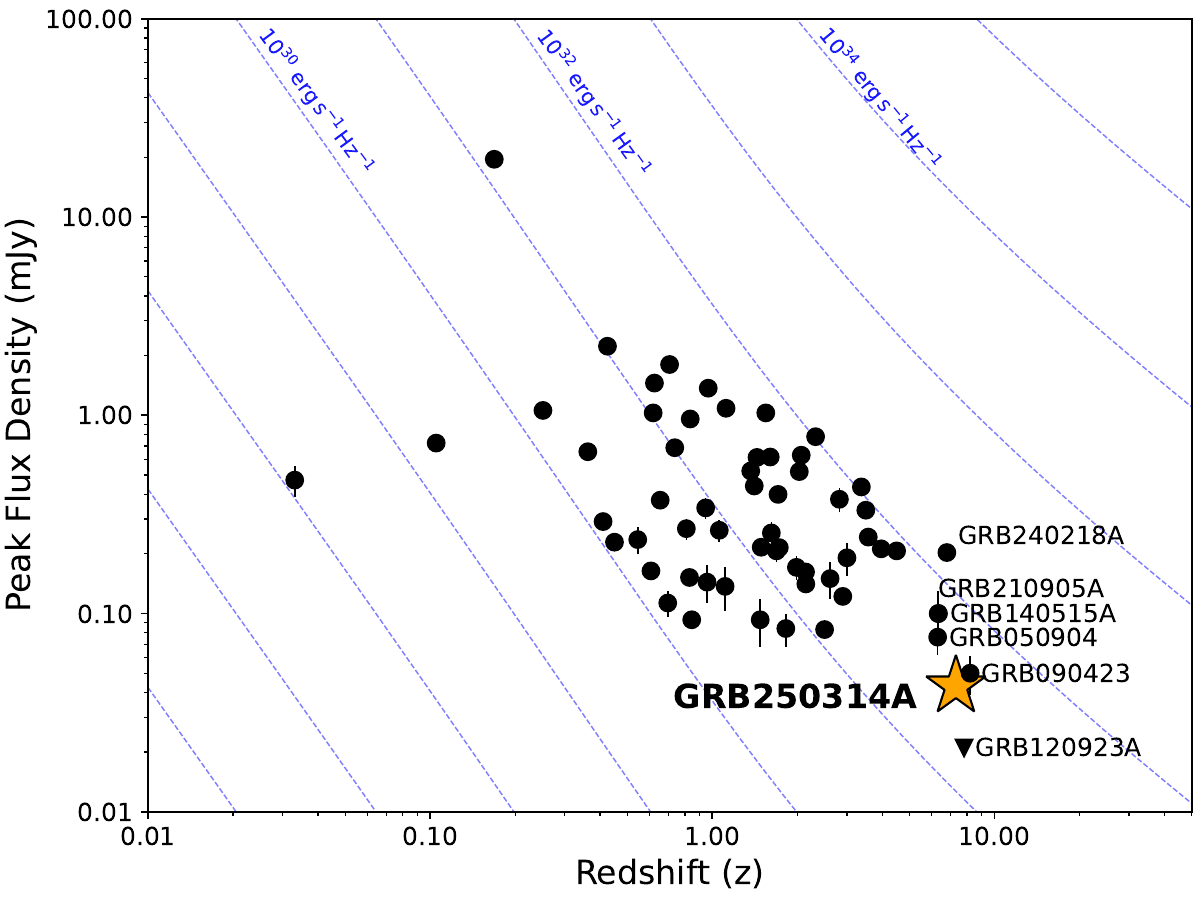}
    \caption{Observed peak flux densities of the radio emission ($\sim8$ GHz) of long GRBs as compared to their redshift. Adapted from \cite{deUgartePostigo2018} using the radio data from \cite{Chandra2012} updated with high-redshift observations \citep{Tanvir2012,rossi:22,brivio:25}. The blue lines represent equal luminosity. Note the overlap between GRB\,210905A and GRB\,140515A data points.}
    \label{fig:radio-img}
\end{figure}

\textit{Interpretation.} 
Figure~\ref{fig:radio-img} shows the peak flux densities of a sample of GRB radio observations. We highlight the observations of GRBs with $z>6$, where GRB\,250314A corresponds to the faintest radio detection, although we note the stronger limit for GRB\,120923A at z$=$7.8. GRB\,250314A has a comparable peak luminosity to the also high-z GRB\,050904 and GRB\,090423 and has a peak luminosity comparable to other lower redshift events.
At about 9~days after the burst (1.1~day rest frame) the  9 GHz detection with ATCA, at a flux consistent with the VLA detection, combined with the tight upper limits at lower frequencies (2.6\,GHz) by MeerKAT indicate a self-absorbed synchrotron spectrum, typical of a reverse-shock component dominating the radio flux at these early times ($\approx1$~day in the rest frame; e.g., \citealt{Frail:2000,Laskar:19}). At later times, when the forward shock is expected to dominate, our radio upper limits are 5-10 times lower that the radio emission observed in a  burst at a similar redshift (GRB\,240218A at $z=6.8$, \citealt{brivio:25}). This may be attributable to either to a lower energy or a more collimated outflow, with the latter indeed consistent with the steep decay observed in the near-infrared, that takes place earlier than observed in GRB\,240218A.

\begin{table}[h]
\caption{Radio observations. Upper limits are given at the  $3\sigma$ level.}
\vspace*{-2ex}

\begin{center}
\begin{tabular}{c c c c}      
\hline\hline
Time & Telescope   & Frequency &  Flux \\
since Tb (days) & & (GHz) & ($\mathrm{\mu Jy}$)\\
\hline

9.42	& MeerKAT  & 1.3 &	<33	\\
29.20	& MeerKAT  & 1.3 &	<24	\\
109.02	& MeerKAT  & 1.3 &	<30	\\
9.29	& MeerKAT  & 2.6 &	<12	\\
29.32	& MeerKAT  & 2.6 &	<14	\\
109.14	& MeerKAT  & 2.6 &	<13.8\\
16.46 & eMERLIN	& 5.1 &	$<90$\\
56.46 & eMERLIN	& 5.1 &	$<125$\\
8.96 &	ATCA & 5.5	& $<33$ \\
6.7     & VLA       &  6  & <60	\\
8.96 &	ATCA &	 9	& $35\pm 10$\\
6.7     & VLA       &  10  & $33\pm 10$	\\
6.7     & VLA       &  15  & $43\pm 12$	\\
\hline
\hline
\end{tabular}
\end{center}
\label{tab:radio}
\end{table}

\section{GRB~250314A Timeline}
\label{appendix:timeline}
\begin{figure*}[h]
    \centering
    \includegraphics[width=0.8\linewidth]
    {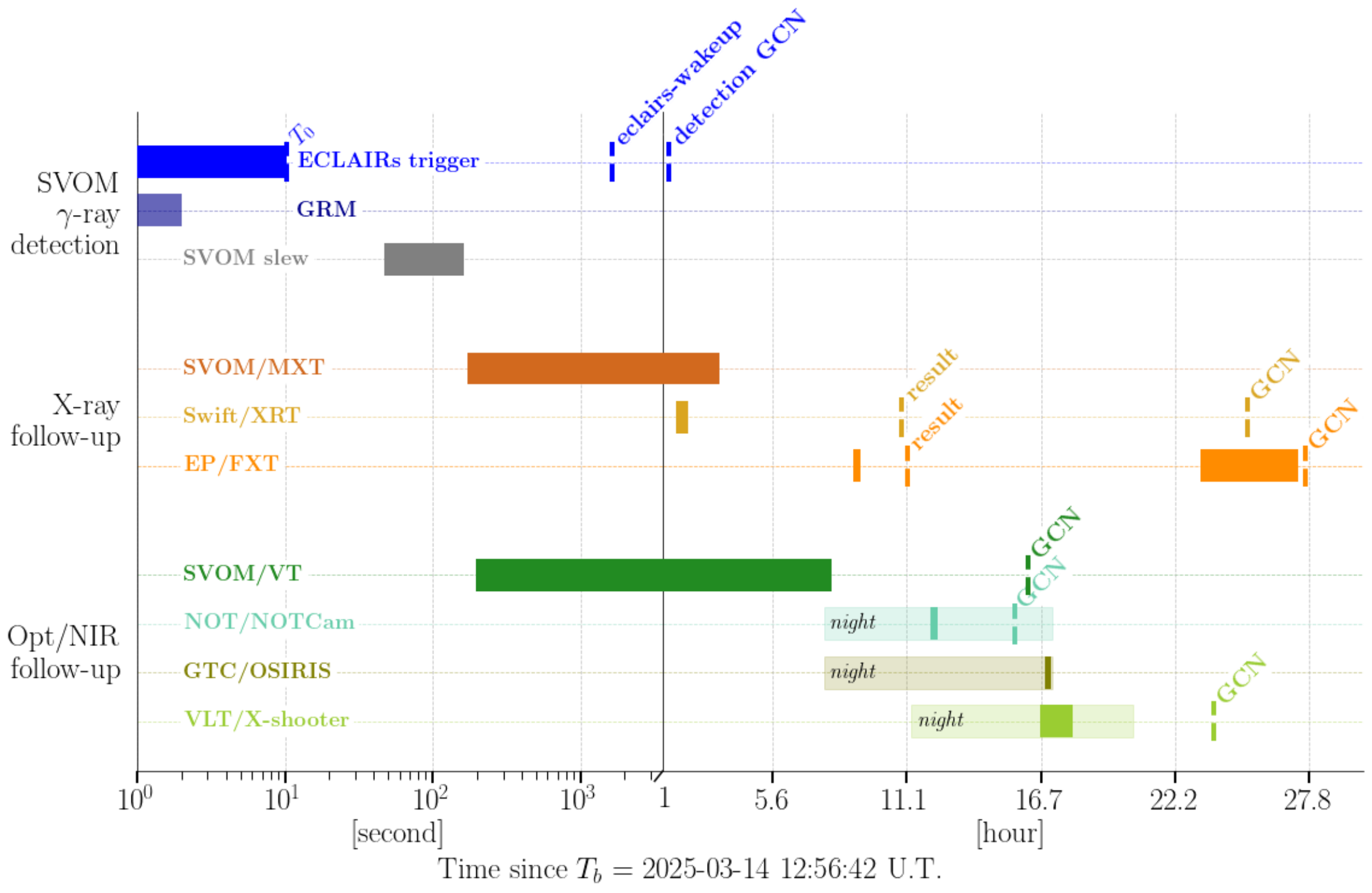}
    \caption{Observation timeline from the \textit{SVOM}/ECLAIRs detection of the burst to the redshift determination by the \textit{VLT}/X-Shooter and the announcement in the GCN Circulars. The color-filled blocks show the observation sequence of the different \textit{SVOM} instruments and follow-up facilities. The shaded blocks for the NOT, GTC and \textit{VLT} telescopes represent the local night time in the Canarians and at Cerro Paranal in Chile. The ``result'' dashed lines for \textit{Swift} and \textit{Einstein Probe} show the time at which the first X-ray source list were issued from the XRT and FXT observations, respectively.}
    \label{fig:timeline}
\end{figure*}

We summarize in Fig.~\ref{fig:timeline} the full observation sequence of GRB~250314A from the ECLAIRs trigger on-board the \textit{SVOM} satellite to the redshift determination at VLT/X-shooter. We distinguish in this timeline the periods of observations and the times at which the corresponding results became available to the community. As discussed in \S~\ref{sec:discussion}, this highlights the importance of setting up efficient coordination and communication between \textit{SVOM} and the follow-up collaborations and clearly shows the delays that we should try to shorten.

\section{GRB\,250314A energetics and classification}
\label{appendix:energetics}

\begin{figure}[h]
\centering
\includegraphics[width=0.9\linewidth]{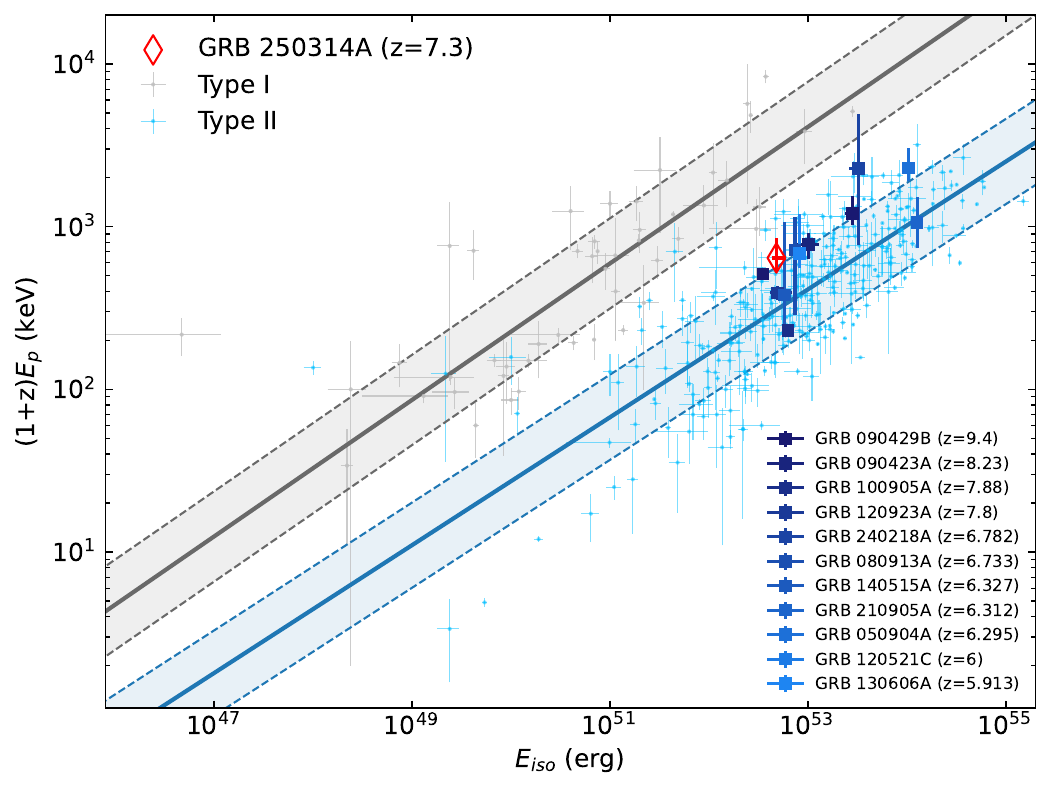}
\caption{GRB\,250314A in the ``Amati diagram'':
The samples of ``short'' (type I, in black) and ``long'' (type II, in blue) GRBs are taken from \citet{lan:23}. The peak energies and isotropic equivalent for the sample of GRBs above $z=6$ are taken from \citet{yasuda:17} (GRB130606A, GRB 120521C, GRB 050904A, GRB140515A, GRB080913A, GRB090423A), \citet{rossi:22} (GRB 210905A), \citet{brivio:25} (GRB 240218A), \citet{Tanvir2018} (GRB 120923A), \citet{gorbovskoy:12} (GRB 100905A) and \citet{Cucchiara2011} (GRB 090429B). All errors were converted to 68\,\% confidence level. The properties of GRB250314A (red) are deduced from the results of the joint ECLAIRs+GRM spectral analysis. 
}
\label{fig:relation}
\end{figure}

Using the results of the time-integrated joint spectral analysis of ECLAIRs+GRM data presented in Sect.~\ref{sec:obs} and Appendix~\ref{appendix:prompt}, and adopting a redshift $z=7.3$ and the corresponding luminosity distance $D_\mathrm{L}=74.0\, \mathrm{Gpc}$  (using $H_0=67.4\, \mathrm{km/s/Mpc}$, $\Omega_\mathrm{m}=0.315$, $\Omega_{\Lambda} = 0.685$, \citealt{Planck_2018}), we deduce a burst 
rest-frame peak energy $(1+z)\,{E}_\mathrm{p}=642^{+209}_{-118}$ keV and an isotropic-equivalent energy $E_\mathrm{iso}=4.65^{+1.13}_{-0.49}\times 10^{52}\, \mathrm{erg}$ in the 10 keV--10 MeV energy range (burst rest frame). This value is well below the probable cutoff around $\sim 4\times 10^{54}\, \mathrm{erg}$ in the distribution of the isotropic equivalent energy \citep{atteia:25}.
We compare GRB\,250314A to other GRBs in the $(1+z)\,E_\mathrm{p}$--$E_\mathrm{iso}$ plane (``Amati relation'', \citealt{amati:02}) in Fig.~\ref{fig:relation}.
GRB\,250314A is clearly located within the region of long GRBs (type~II).

\section*{Acknowledgments}

The Space-based multi-band Variable Objects Monitor (\textit{SVOM}) is a joint Chinese French mission led by the Chinese National Space Administration (CNSA), the French Space Agency (CNES), and the Chinese Academy of Sciences (CAS). We gratefully acknowledge the unwavering support of NSSC, IAMCAS, XIOPM, NAOC, IHEP, CNES, CEA, CNRS, University of Leicester, and MPE.
We acknowledge the observational data taken at VLT (program 114.27PZ, PIs: Tanvir, Vergani, Malesani), NOT (program P71-506, PIs: Malesani, Fynbo, Xu), GTC (program GTCMULTIPLE4G-25A, PI:Ag\"u\'i Fern\'andez).

This work also makes use of data obtained with the Einstein Probe, a space mission supported by the Strategic Priority Program on Space Science of the Chinese Academy of Sciences, in collaboration with ESA, MPE and CNES (grant XDA15310000).
We acknowledge the use of public data from the Swift data archive.
The Australia Telescope Compact Array is part of the Australia Telescope National Facility (https://ror.org/05qajvd42) which is funded by the Australian Government for operation as a National Facility managed by CSIRO. We acknowledge the Gomeroi people as the Traditional Owners of the Observatory site.
e-MERLIN is a National Facility operated by the University of Manchester at Jodrell Bank Observatory on behalf of STFC, part of UK Research and Innovation.
The MeerKAT telescope is operated by the South African Radio Astronomy Observatory, which is a facility of the National Research Foundation, an agency of the Department of Science and Innovation.
\\
\noindent
AS acknowledges support by a postdoctoral fellowship from the CNES.
SDV and BS acknowledge the support of the French Agence Nationale de la Recherche (ANR), under grant ANR-23-CE31-0011 (project PEGaSUS).
LP, ALT  GB and GG, acknowledge support by the European Union horizon 2020 programme under the AHEAD2020 project (grant agreement number 871158). LP, ALT and GG also acknowledge  support by ASI (Italian Space Agency) through the Contract no. 2019-27-HH.0.
NRT acknowledges support from STFC grant ST/W000857/1.
The authors are thankful for support from the National Key R\&D Program of China (grant Nos. 2024YFA1611704,2024YFA1611700). This work is supported by the Strategic Priority Research Program of the Chinese Academy of Sciences (Grant No.XDB0550401), and by the National Natural Science Foundation of China (grant Nos. 12494575, 12494571, 12494570 and 12133003).

\end{appendix}
\end{document}